\documentclass[prd,nofootinbib,showpacs,preprint]{revtex4}
\usepackage{graphicx}
\usepackage{color}
\usepackage{epsfig}
\usepackage{amsmath}
\usepackage{amsfonts}
\usepackage{amssymb}
\usepackage{url}
\usepackage{hyperref}
\usepackage{subfigure}
\usepackage{cancel}
\newcommand{\beqa}{\begin{eqnarray}}
\newcommand{\eeqa}{\end{eqnarray}}
\newcommand{\beq}{\begin{equation}}
\newcommand{\eeq}{\end{equation}}

\newcommand{\nn}{\nonumber}
\newcommand{\bmt}{\begin{pmatrix}}
\newcommand{\emt}{\end{pmatrix}}
\usepackage[toc,page]{appendix}
\usepackage{comment}
\usepackage[sort&compress]{natbib}
\newcommand{\be}{\begin{equation}}
\newcommand{\ee}{\end{equation}}
\newcommand{\bea}{\begin{eqnarray}}
\newcommand{\eea}{\end{eqnarray}}

\usepackage[normalem]{ulem}

\newcommand{\nua}[1]{\ensuremath{\rlap{\kern-2.5pt\ensuremath{\overset{\scriptscriptstyle(-)}{\phantom{\nu}}}}{\ensuremath{{\nu}_{#1}}}}\xspace}

\newcommand{\eVq}  {\text{eV}^2}

\begin{document}
\title{Scalar triplet leptogenesis in the presence of right-handed neutrinos with $S_3$ symmetry}
\author{Subhasmita Mishra and Anjan Giri}
\affiliation{Department of Physics, IIT Hyderabad, Kandi-502285,  India  }

\begin{abstract}

Leptogenesis appears to be a viable alternative to account for the baryon asymmetry of the Universe through baryogenesis. In this context, we consider a scenario in which the standard model is extended with $S_3$ and  $Z_2$ symmetry in addition to the two scalar triplets, two scalar doublets and three right handed neutrinos.  Presence of scalar triplets and right-handed neutrinos in the scenarios of both type-I and type-II seesaw framework provide a different leptogenesis option and can help us to understand the matter-antimatter asymmetry with simple $S_3$ symmetry. We discuss the neutrino phenomenology and leptogenesis in both high ($O(10^{10})$ GeV)  and low energy scale (O(2)TeV) by constraining the Yukawa couplings. Moreover, we also consider the constraints on model parameters from neutrino oscillation data and leptogenesis to explain the rare lepton flavor violating decay and muon g-2 anomaly. 

\end{abstract}

\maketitle
\section{I\lowercase{ntroduction}}

The Standard Model (SM) of particle physics has attained an unprecedented level of success over the last few decades, which culminated with the discovery of the Higgs boson at the CERN Large Hadron Collider. However, there appears to be observations which cannot be explained within the framework of the SM. In this context, the observation of neutrino oscillation has indicated that the SM needs to be extended to accommodate the massive neutrinos \cite{pdg:2016}. Moreover, there exists evidence of the baryon asymmetry of the universe, with the obtained value of $\Omega_B h^2 = 0.0223\pm 0.0002$ \cite{PLANCK:2016} that corresponds to the baryon asymmetry $Y_B\equiv {\eta_B/s} \approx 0.86 \times 10 ^{-10}$. There have been many attempts to find some hint of the physics beyond the SM (BSM) but the quest so far remains unsuccessful. In the absence of any clear cut idea so as to ascertain the nature of the new physics it is quite natural to explore simple extensions of the SM which can help to explain the observed data.

The leptonic sector, in particular the study of neutrinos has taken the center stage in particle physics in recent years. Discrete symmetries are widely used for a long time for BSM model building and to explain the neutrino phenomenology \cite{King:2015bja}-\cite{DeRujula:1977dmn}. The discrete symmetries commonly discussed are the $S_3$, $S_4$ and $A_4$ symmetries to explain the observed neutrino oscillation data. Here we choose the simplest permutation symmetry, the $S_3$ along with $Z_2$ symmetry, to explain neutrino mass and also discuss leptogenesis \cite{Kubo:2003iw}- \cite{Meloni:2010aw}. In addition to the SM particle spectrum, we introduce three right handed neutrinos, two Higgs doublets and two scalar triplets to explain the neutrino mass with type I+ II  seesaw mechanism \cite{Hambye:2005tk}. There are a lot of studies using $S_3$ symmetry but here we would like to add another aspect of it to the growing list of possibilities. Earlier, it has been discussed in the literature that $S_3$ symmetry with type-I seesaw scenario could be helpful to accommodate the experimental findings in both quark and lepton sectors, in addition to explaining leptogenesis. Despite the simplicity of the type I seesaw model, type II seesaw is equally frequented due to the fact that addition of scalar does not lead to any anomaly, neither does it have negative contribution to the radiative correction of the SM Higgs mass, unlike the fermions. Leptogensis with $S_3$ symmetry and right handed neutrinos have been considered before in \cite{Araki:2005ec} and here we consider the possible effect of scalar triplets with $S_3$ symmetry.
       
 The CP violation prescribed by the Kobayashi-Maskawa mechanism of the SM is not capable of explaining the observed matter antimatter asymmetry of the Universe and, therefore, lepton asymmetry plays a significant role here. Leptogenesis appears to be an elegant mechanism where the asymmetry generated in the leptonic sector of the SM can be converted to baryogenesis through sphaleron transitions and, in fact, this idea looks to be very promising. In general, lepton asymmetry produced by the out of equilibrium decay of right handed neutrinos has been widely studied in the literature \cite{Grimus:2003sf}-\cite{Abada:2006ea}. But there are very few studies devoted to the generation of lepton asymmetry through the out of equilibrium decay of the scalar triplets in type II seesaw framework \cite{Sierra:2014tqa}-\cite{Ma:1998dx}. Nonzero CP asymmetry cannot be generated with one loop contribution in the presence of only one scalar triplet \cite{Felipe:2013kk}. Hence scalar sector should be extended with at least one more triplet to generate a nonzero CP asymmetry from the interference of tree and one loop contributions. Since the scalar triplet has two different decay modes, even though the gauge interactions and the total decay rate of the triplets are larger than the expansion rate of the universe, still the lepton asymmetry can be generated with any of the decays being out of equilibrium. There exist studies in the literature in connection with the leptogenesis from scalar triplet in the presence of right handed neutrinos \cite{AristizabalSierra:2012pv},\cite{Hambye:2005tk}, and we focus here in this direction with some additional symmetries.
         
 Motivated by the need to look for scenarios beyond the SM, we chose the simplest discrete symmetry ($S_3$) with minimal particle contents to discuss neutrino phenomenology and leptogenesis as a viable option. Generation of lepton asymmetry from the decay of heavy Majorana fermions is well described in the literature in the framework of  $S_3$ symmetry. But few studies have done with scalar triplets in this context. Therefore, we tried to explain the leptogenesis phenomena from the decay of heavy scalar triplet and its interesting phenomenology in the presence of right-handed Majorana neutrinos in a composite seesaw (type I+II) scenario.
   
 The remainder of the article is as follows: in Section II, we explain the model framework. Here we describe the Lagrangian with $S_3$ symmetry including the extended particle contents. Extended Higgs sector is also discussed in this section. Section III includes the description of the neutrino masses and mixing with type I $+$ II seesaw mechanism, where we discuss the relevant framework to compare our results with the observed data. Section IV is devoted to the leptognesis. Here we discuss all the possible scenarios and associated Boltzmann equation including the results.  Section V contains the Conclusion.
 
\section{The model}
In this section, we discuss the particle content and corresponding group charges of the SM and extra particles, excluding the quark sector and focus only on the leptonic sector. The importance of discrete symmetries in particle phenomenology has already been discussed extensively in various studies earlier \cite{King:2013eh}, \cite{Mondragon:2006hi}.  We consider the extension of the SM (${\rm SU}(3)\times \rm SU(2)_L\times \rm U(1)_Y)$) with the simplest non-abelian discrete flavor symmetry, $S_3$, and the abelian symmetry $Z_2$. In addition to the SM particles, we include three right handed neutrinos($ N_{(1,2,3)R}$),  two Higgs doublets, and two Higgs triplets ($\Delta_{1,2}$) to explain the neutrino mixing and leptogenesis. 

\begin{table}[h]
\begin{center}
\begin{tabular}{| c | c | c | c | }
\hline
~Particles~ & $\rm SU(3)_c \otimes SU(2)_L\otimes U(1)_Y$ ~ &~ $S_3$ ~& ~$Z_2$ \\
\hline
\hline
 $L_e, L_\mu$ & ($1,2,-1$) & $2$ & $+1$ \\
  $L_\tau$ & ($1,2,-1$) & $1$ & $+1$ \\
  $E_{1R}, E_{2R}$ & ($1,1,-2$) & $2$ & $+1$ \\
  $E_{3R}$ & ($1,1,-2$) & $1$ & $+1$ \\
   $N_{1R}, N_{2R}$ & ($1,1,0$) & $2$ & $+1$\\
   $N_{3R}$ & ($1,1,0$) & $1$ & $-1$ \\ 
   $H_1,H_2$ & ($0,2,1$) & $2$ & $+1$\\
   $H_3$ & ($0,2,1$)& $1$ & $-1$ \\
   $\Delta_1$ & $(0,3,2)$ &$1$ & $+1$\\
   $\Delta_2$ & $(0,3,2)$ &$1$ & $+1$\\
   \hline
\end{tabular}
\end{center}
\caption{Particle contents and quantum numbers under SM, $S_3$, and $Z_2$ }
\label{particle content table}
\end{table}
 In Table~\ref{particle content table}, $L_e$, $L_\mu$ and $L_\tau$  are the first, second and third generation lepton families respectively, $N_{iR}$ and $\Delta_{1,2}$ are the right handed singlet Majorana neutrinos and $\rm SU(2)$ triplet Higgs, respectively. The scalar triplets are defined in $\rm SU(2)$ basis and is given by \cite{Sierra:2014tqa}
\begin{eqnarray}
\Delta_i=\begin{pmatrix}
\frac{{\Delta_i}^+}{2} && {\Delta_i}^{++}\\
{\Delta_i}^0 && -\frac{{\Delta_i}^+}{2}\\
\end{pmatrix}.
\end{eqnarray}
The invariant Lagrangian for both type I and type II, involving the scalars and fermions in the framework under consideration ($\rm SU(3)_c \otimes SU(2)_L\otimes U(1)_Y \otimes S_3\otimes Z_2$), is given by \cite{Araki:2005ec}\cite{Akhmedov:2006de}
\begin{eqnarray}
\mathcal{L} &\supset & y_{t1}\left[\overline{\tilde{L_e}} \Delta_1 L_e+\overline{\tilde{L_\mu}} \Delta_1 L_\mu \right] +{y_{t1}}'[\overline{\tilde{L_\tau}}\Delta_1 L_\tau] \nonumber\\
&&+ y_{t2}\left[\overline{\tilde{L_e}}\Delta_2 L_e+\overline{\tilde{L_\mu}}\Delta_2 L_\mu \right] +{y_{t2}}'[\overline{\tilde{L_\tau}}\Delta_2 L_\tau] \nonumber \\
&& -y_{\nu_1}\left[ \overline{L_e} \tilde{H_2} N_{1R} +\overline{L_\mu} \tilde{H_1} N_{1R}+\overline{L_e} 
\tilde{H_1} N_{2R}-\overline{L_\mu} \tilde{H_2} N_{2R} \right] \nonumber\\
               && - y_{\nu_3}\left[ \, \overline{L_\tau}  \,  \,  
         \tilde{H_1} N_{1R}+\, \overline{L_\tau}  \, \,  
         \tilde{H_2}  N_{2R}\right]- y_{\nu_4}\,\left[\overline{L_\tau} \,  \, 
         \tilde{H_3} N_{3R} \right] \nonumber \\
&&- y_{l2} \left[\overline{L_e} {H_2} E_{1R} +\overline{L_\mu} {H_1} E_{1R}+
        \overline{L_e}  {H_1} E_{2R}-\overline{L_\mu} {H_2} E_{2R}\right] \nonumber \\
&&- y_{l4}\left[ \, \overline{L_\tau}  \, \,  {H_1}  E_{1R}+\overline{L_\tau}  \,  \,  {H_2} E_{2R}\right]- 
y_{l5} \, [\overline{L_e} \, \,{H_1}  E_{3R}+\overline{L_\mu} \,  \, {H_2} E_{3R}]\nonumber\\
&& -\frac{1}{2} \sum_{i=1,2}\overline{N}_\text{iR}^\text{c} M_\text{iR} N_\text{iR} -\frac{1}{2}\overline{N}_\text{3R}^\text{c} M_\text{3R} N_\text{3R}+
	\rm{~h.c} - \rm V(H_i,\Delta_j)\hspace{2mm} (i=1,2,3;  j=1,2).
\label{Model interaction lagrangian}
\end{eqnarray}
In the above expression, $\overline{\tilde{L_L}}=\overline{{L^c}_L} i\tau_2=(-\overline{{e^c}_L} \hspace{3mm}   \overline{{\nu^c}_L} )$, $y_{\nu_i}$ and $y_{li}$ are the Yukawa couplings of neutral and charged leptons, respectively. $M_{iR}$ are the Majorana masses of right handed neutrinos. Models with extra Higgs in the presence of discrete symmetries are well studied in the literature \cite{Branco:2006ce}-\cite{Koide:2005ep}. With the additional scalar content in the model, we can write the interaction potential as
\begin{eqnarray}
V(H_i, \Delta_j)&=&m^2_0 {H_3}^{\dagger}H_3+m^2_d({H_2}^{\dagger}H_2+{H_1}^{\dagger}H_1)+\lambda_1({H_2}^{\dagger}H_2+{H_1}^{\dagger}H_1)^2 \nn \\
&&+\lambda_2({H_1}^\dagger H_2-{H_2}^\dagger H_1)^2+\lambda_3[({H_1}^\dagger H_1-{H_2}^\dagger H_2)^2+({H_1}^\dagger H_2+{H_2}^\dagger H_1)^2] \nn \\
&&+\lambda_4[({H_3}^\dagger H_1)({H_1}^\dagger H_2+{H_2}^\dagger H_1)+({H_3}^\dagger H_2)({H_1}^\dagger H_1-{H_2}^\dagger H_2)+h.c] \nn \\
&&+\lambda_5[({H_3}^\dagger H_3)({H_1}^\dagger H_1+{H_2}^\dagger H_2)]+\lambda_6[({H_3}^\dagger H_1)({H_1}^\dagger H_3)+({H_3}^\dagger H_2)({H_2}^\dagger H_3)] \nn \\
&&+\lambda_7[({H_3}^\dagger H_1)({H_3}^\dagger H_1)+({H_3}^\dagger H_2)({H_3}^\dagger H_2)+h.c]+\lambda_8({H_3}^\dagger H_3)^2 \nn \\
&&+\mu^2_{SB1}({H_1}^\dagger H_2+h.c)+\mu^2_{SB2}({H_3}^\dagger(H_1+H_2)) \nn \\
&&+m^2_{t1} {\rm Tr}({\Delta_1}^{\dagger} \Delta_1)+m^2_{t2} {\rm Tr}({\Delta_2}^\dagger \Delta_2)+\mu_1({\tilde{H_1}}^\dagger {\Delta_1}^\dagger H_1+{\tilde{H_2}}^\dagger {\Delta_1}^\dagger H_2) \nonumber \\
&&+{\mu_1}'({\tilde{H_3}}^\dagger {\Delta_1}^\dagger H_3)+\mu_2({\tilde{H_1}}^\dagger {\Delta_2}^\dagger H_1)+{\mu_2}'({\tilde{H_3}}^\dagger {\Delta_2}^\dagger H_3)+g_2({H_3}^{\dagger} {\Delta_1}^{\dagger} \Delta_1 H_3)\nonumber \\
&&+g_1( {H_1}^{\dagger} {\Delta_1}^{\dagger} \Delta_1 H_1+{H_2}^{\dagger} {\Delta_1}^{\dagger} \Delta_1 H_2)+g_3({H_1}^{\dagger} {\Delta_2}^{\dagger} \Delta_2 H_1)+ g_4({H_3}^{\dagger} {\Delta_2}^{\dagger} \Delta_2 H_3) \nonumber \\
&&+k_1(({H_1}^\dagger H_1){\rm Tr}({\Delta_1}^\dagger {\Delta_1})+({H_2}^\dagger H_2) {\rm Tr}({\Delta_1}^\dagger {\Delta_1}))+ k_2(({H_3}^\dagger H_3){\rm Tr}({\Delta_1}^\dagger {\Delta_1}))\nonumber \\
&&+k_3(({H_1}^\dagger H_1) {\rm Tr}({\Delta_2}^\dagger {\Delta_2})+({H_2}^\dagger H_2) {\rm Tr}({\Delta_2}^\dagger {\Delta_2})) +k_4(({H_3}^\dagger H_3) {\rm Tr}({\Delta_2}^\dagger {\Delta_2})) \nn \\
&& +t_1 {\rm Tr}({\Delta_1}^\dagger \Delta_1)^2+t_2 {\rm Tr}({\Delta_2}^\dagger \Delta_2)^2+t_3 {\rm Tr}({\Delta_1}^\dagger \Delta_1) {\rm Tr}({\Delta_2}^\dagger \Delta_2).
\label{scalar potential}
\end{eqnarray}
The minimization conditions are given by $\frac{\partial V}{\partial v_1}=0$, $\frac{\partial V}{\partial v_3}=0$, $\frac{\partial V}{\partial u_1}=0$, $\frac{\partial V}{\partial u_2}=0$, where, $\langle {\Delta_{1,2}}\rangle=\begin{pmatrix} 0 & 0 \\ u_{1,2} & 0 \end{pmatrix}$, $\langle H_1 \rangle=\begin{pmatrix} 0 \\ v_1 \end{pmatrix}$,  $\langle H_2 \rangle=\begin{pmatrix} 0 \\ v_2 \end{pmatrix}$ and $\langle H_3 \rangle=\begin{pmatrix} 0 \\ v_3 \end{pmatrix}$.\\ 
We found the stability conditions of the scalar potential by using the co-positivity criteria \cite{Kannike:2012pe}, which are given below  
\begin{eqnarray}
&& \lambda_1+\lambda_3\geq 0, \hspace{2mm} \lambda_8\geq 0,l_1\geq 0,l_2 \geq 0, \nn \\
&& \lambda_5+\lambda_6+|\lambda_7|+\sqrt{\lambda_8(\lambda_3+\lambda_1)}\geq 0, \nn \\
&& 3(\lambda_1+\lambda_3)\sqrt{\lambda_8}+2(\lambda_5+\lambda_6+|\lambda_7|)\sqrt{\lambda_1+\lambda_3}\geq 0, \nn  \\
&& 2(\lambda_5+\lambda_6+|\lambda_7|)^2-3\lambda_8(\lambda_1+\lambda_3)\geq 0,  \nn \\
&& (g_1+k_1)+\sqrt{(\lambda_1+\lambda_3)t_1}\geq 0,\hspace{4mm}(g_3+k_3)+\sqrt{(\lambda_1+\lambda_3)t_2}\geq 0, \nn \\
&& (g_2+k_2)+\sqrt{\lambda_8 t_1}\geq 0, \hspace{3mm}(g_4+k_4)+\sqrt{\lambda_8 t_2}\geq 0,\hspace{3mm}t_3+\sqrt{t_2 t_1}\geq 0.
\label{stability}
\end{eqnarray}
The explicit symmetry breaking terms $\mu_{SB1}$ and $\mu_{SB2}$ in the potential in Eq.\eqref{scalar potential} break $S_3\times Z_2$ softly  to another symmetry, $H_1\leftrightarrow H_2$. Hence one can choose, $\langle H^0_1 \rangle=\langle H^0_2 \rangle=v_1$. The symmetry breaking terms are important for generating masses of additional Higgs particles. Multi Higgs doublet models allow tree level Flavor Changing Neutral Current (FCNC), unless the coupling of scalar doublets to both up and down type quarks and leptons are protected. These FCNCs can be suppressed with Higgs masses at TeV scale, which cannot be generated by electroweak symmetry breaking. Therefore, one can adjust the heavy Higgs mass by fine-tuning the soft breaking parameters.
\subsection{Masses and Mixing in the Higgs sector}
Looking at the symmetry breaking terms in the scalar potential in Eq.\eqref{scalar potential} one can redefine $H_1$ and $H_2$  in terms of $H_+$ and $H_-$ as \cite{Araki:2005ec}, \begin{eqnarray}
H_1=\frac{H_+ + H_-}{\sqrt{2}}, \hspace{3mm}
H_2=\frac{H_+ - H_-}{\sqrt{2}}.
\label{redefhiggs}
\end{eqnarray}
After redefinition, we can have $\langle H^0_- \rangle =0$ and $\langle H^0_+ \rangle = v_+$, with the assumption that $v_1=v_2$. The mixing between $H_+$ and $H_3$ can be considered as both of them acquire a non zero Vacuum Expectation Value (VEV). Here we can write the mass basis of these two Higgs fields by orthogonal rotation of flavor states as follows,
\begin{eqnarray}
\begin{pmatrix}
H_3\\
H_+\\
\end{pmatrix}=\begin{pmatrix}
\cos{\beta} & \sin{\beta}\\
-\sin{\beta} & \cos{\beta}\\
\end{pmatrix} \begin{pmatrix}
H_L\\
H_H\\
\end{pmatrix},
\label{higgsmixing}
\end{eqnarray}
where, $H_L=H_3 \cos{\beta}-H_+ \sin{\beta}$ and $H_H=H_3 \sin{\beta} + H_+ \cos{\beta} $, and $\beta$ is the Higgs mixing angle. The new Higgs fields are written in $SU(2)$ doublet form as  
\begin{eqnarray}
H_L=\begin{pmatrix}
0\\
h^0_L+v
\end{pmatrix},~~~  H_H=\begin{pmatrix}
h^+_H\\
h^0_H+ia_H\\
\end{pmatrix},~~~  H_-=\begin{pmatrix}
h^+_-\\
h^0_-+ia_-
\end{pmatrix},
\end{eqnarray}
where, $H_L$ is the SM like Higgs with VEV, $v=\sqrt{{v_+}^2+{v_3}^2}=246$ GeV and $\tan{\beta}=\frac{v_+}{v_3}$ with $\langle H^0_3 \rangle=v_3$. Charged and CP odd components of $H_L$ will be absorbed by the SM gauge bosons to acquire mass in unitary gauge conditions. And rest of the Higgs doublets will have two CP odd, two charged and two neutral scalar fields.\\
The masses of the scalar fields are given by
\begin{eqnarray*}
&&M^2_{h_H}\approx M^2_{{h_H}^+} \approx M^2_{a_H} \approx \mu^2_{SB1} \cos^2{\beta}+2 \sqrt{2} \mu^2_{SB2}\sin{\beta}\cos{\beta}+(\mu_1 u_1+\mu_2 u_2),\\
&& M^2_{h_L}\approx \mathcal{O} (v^2),\\
&& M^2_{h_-}\approx M^2_{{h_-}^+}\approx M^2_{a_-}\approx \mu^2_{SB1}+\mu_1 u_1+\mu_2 u_2 .
\end{eqnarray*}
The masses of the Higgs fields, other than the SM Higgs (125 GeV), can be achieved to be order of TeV by finetuning, which help in suppressing the tree level FCNCs. 
\subsection*{Phase Re-absorption}
The phases of complex fermion fields can be redefined by fixing the phases in the complex Yukawa coupling present in the Lagrangian in Eq.\eqref{Model interaction lagrangian} \cite{Araki:2005ec}. Let's consider that the neutral lepton Yukawa couplings transform as $y_{\nu_i}\rightarrow e^{i p_{yi}} y_{\nu_i}$~($i=1,3,4$), where, $p_{yi}$ are the phases of transformations. Similarly for the charged lepton Yukawa and the fermion fields transform as $y_{li}\rightarrow e^{i p_{yli}} y_{li}$~($i=2,4,5$), $L_i\rightarrow e^{ip_l} L_i$~($i=1,2$), $L_3 \rightarrow e^{i p_{l3}}L_3, E_{iR}\rightarrow e^{ip_E} E_{iR}$~($i=1,2$), $E_{3R}\rightarrow e^{i p_{E3}} E_{3R}, N_{iR}\rightarrow e^{ip_R} N_{iR}$~($i=1,2$) and  $ N_{3R} \rightarrow e^{i p_{3R}} N_{3R} $.

Phases of $N_{iR}$ can be absorbed in the Majorana mass matrix $M_{iR}$,  $M_{3R}$ and the phases in the charged lepton Yukawa couplings can be fixed by the redefinition of the fermion fields, which can be found from the Lagrangian in Eq.\eqref{Model interaction lagrangian} as follows
\begin{eqnarray}
&& p_l=-p_{y_{l2}}+p_E, \hspace{3mm} p_{l3}=-p_{y_{l4}}+p_E, \hspace{3mm} p_{E3}=p_{y_{l5}}-p_{y_{l2}}+p_E.
\end{eqnarray}
Hence one can choose the leftover phase in charged lepton sector to be $p_l$, which can be fixed by the Yukawa coupling of the neutrinos, i.e.,  $p_l=-p_{y1}$. Therefore, the remnant phase in the neutral lepton complex Yukawa coupling are $p_{y3}$ and $p_{y4}$. While constructing the  mass matrices, the neutral and charged lepton fields can be rotated separately, hence only the relative phase $p_{y3}$ - $p_{y4}$ appears in the neutrino mass matrix.  

Similarly, in the scalar sector  the triplet lepton Yukawa and the complex triplet fields transform as $y_{ti}\rightarrow e^{i p_{y_{ti}}} y_{ti}$ and $\Delta_i \rightarrow e^{i p_{\Delta_i}} \Delta_i$,  respectively. From the triplet Lagrangian in Eq.\eqref{Model interaction lagrangian} one can fix the phases as, $p_{y_{t1}}=2 p_l+p_{\Delta_1}$ and $p_{{y_{t1}}'}=2 p_{l3}+p_{\Delta_1}$. Hence, if the phases of $y_{ti}$ can be absorbed by the redefinition of $\Delta_i$, the remnant phases in the scalar triplet-lepton interaction sector are $p_{y_{ti'}}$.

\section{Neutrino Masses and Mixing}

In order to discuss the neutrino masses and mixing, we first discuss the type I seesaw mass matrix for neutral leptons, which is given in the basis of 
$\tilde{N} = (\nu_\text{L}^\text{c}, ~N_\text{R})^\text{T}$ as 
\begin{equation}
	\mathcal{M} = 
	\begin{pmatrix}
		0 & M_\text{D} \\
		M_\text{D}^\text{T} & M_R  \\
	\end{pmatrix}.
\end{equation}
We consider the light neutrino mass formula, which is described by the well known type I seesaw mechanism as \cite{Mohapatra:2006gs},\cite{Miranda:2016ptb}
\begin{align}
	\mathcal{M^I_\nu} &= M_\text{D} M_{R}^{-1} \left( M_\text{D}  \right)^T.
	\label{type I formula}
\end{align}
Looking at the Lagrangian in Eq.\eqref{Model interaction lagrangian}, one can write the flavor structure of Dirac mass matrix for the neutral and charged leptons as
\begin{eqnarray}
M_D=\begin{pmatrix}
   m_1 & m_1      & 0 \\
   m_1  & -m_1 & 0 \\
   m_3       & m_3      & m_4 \rm e^{i \phi}\\
   \end{pmatrix} \hspace{2mm},\hspace{2mm}
M_l=\begin{pmatrix}
   m_{l2}       & m_{l2}       & m_{l5}\\
   m_{l2}       & -m_{l2}  & m_{l5} \\
   m_{l4}       & m_{l4}   & 0 
\\
   \end{pmatrix},
   \label{mass matrices} 
\end{eqnarray}\\
where, $m_1=y_{\nu_1} v_1$, $m_3=y_{\nu_3} v_1$, and $m_4= y_{\nu_4} v_3$. Similarly, $m_{l2}=y_{l2} v_1$, $m_{l5}=y_{l5} v_1$, and $m_{l4}=y_{l4} v_1$ with the assumption that $v_1=v_2$. Using the mixing of the Higgs fields and the redefinition of VEVs, one can rewrite the terms inside the Dirac mass matrix as,  $m_1=y_{\nu_1} v \sin{\beta}$, $m_3=y_{\nu_3} v \sin{\beta}$, and $m_4= y_{\nu_4} v \cos{\beta}$. Similarly, one can also write, $m_{l2}=y_{l2} v \sin{\beta}$, $m_{l5}=y_{l5} v\sin{\beta}$, and $m_{l4}=y_{l4} v \sin{\beta}$. From the phase re-absorption, as explained before, we can always have the choice to put the remnant phase in any element of Dirac term. Here we put the relative phase ($\phi=p_{y3}-p_{y4}$) in  $m_4$ for simplicity in calculation. Similarly, the type II mass matrix can be constructed from the Lagrangian in Eq.\eqref{Model interaction lagrangian} by using the general formula of effective neutrino mass matrix in type II seesaw mechanism \cite{Akhmedov:1999tm}, i.e $\mathcal{M^{II}_\nu}= \sum_i  \frac{2   \mu_{iL} Y_{\Delta_i} v^2}{M^2_{\Delta_i}}$. The structure of mass matrix is given below 
\begin{eqnarray}
\mathcal{M^{II}_\nu}=\begin{pmatrix}
x_1 y_{t1}+x_2 y_{t2} &0 &0\\
0& x_1 y_{t1}+x_2 y_{t2} &0\\
0& 0& x_1 {y_{t1}}'+x_2{y_{t2}}'
\end{pmatrix},
\label{type II matrix}
\end{eqnarray}\\
where, $x_i= \frac{2 \mu_{iL} v^2}{M^2_{\Delta_i}} (i=1,2)$ and $ \mu_{iL}=2\mu_i \sin^2{\beta}+{\mu_i}' \cos^2{\beta}$. Here $\mu_{iL}$ are the coupling of triplets to the SM like Higgs($H_L$) in this model, which contribute to the neutrino mass.

\subsection{Diagonalization of charged Lepton and Neutrino mass matrices }
The squared charged lepton mass matrix can be diagonalized by unitary transformation as $U_{eL} M_l {M_l}^\dagger {U_{eL}}^\dagger={\rm Diag}(m^2_e,\hspace{2mm}m^2_{\mu},\hspace{2mm} m^2_\tau)$. The Eigenvector matrix of the squared masses can be obtained by solving the characteristic equation \cite{Mondragon:2006hi}, 
\begin{eqnarray}
U_{el}=\begin{pmatrix}
   \frac{x}{\sqrt{2(1-x^2)}} && \frac{1}{\sqrt{2(1+x^2)}} && \frac{1}{\sqrt{2(1+\sqrt{z})}}\\
  \frac{-x}{\sqrt{2(1-x^2)}}  &&\frac{-1}{\sqrt{2(1+x^2)}}  && \frac{1}{\sqrt{2(1+\sqrt{z})}}\\
   \frac{\sqrt{1-2x^2}}{\sqrt{1-x^2}} && \frac{x}{\sqrt{1+x^2}} && \frac{\sqrt{z}}{\sqrt{(1+\sqrt{z})}}
   \end{pmatrix}.
   \label{lepton mixing matrix}
\end{eqnarray} 
Where, $x=\frac{m_e}{m_\mu}$, and $z= \frac{m_e^2m_\mu^2}{m_\tau^4}$. With the consideration of the Majorana neutrinos to be in diagonal basis, for simplicity, we can write the effective small neutrino mass matrix $\mathcal{M^I_\nu}$ in type I seesaw framework from Eq.\eqref{type I formula} and \eqref{mass matrices} as 
 \begin{eqnarray}
      \mathcal{ M^I_\nu} &=& \begin{pmatrix}
    {\eta_1}^2 & 0 & \eta_1 \eta_3\\
   0 & {\eta_1}^2 & 0\\
   \eta_1 \eta_3 & 0 & {\eta_3}^2+{\eta_4}^2 e^{2i\phi}
   \end{pmatrix},
   \label{typeImass}
 \end{eqnarray}
where, $\eta_1=\frac{\sqrt{2} m_1}{\sqrt{M_{1R}}}, \eta_3=\frac{\sqrt{2} m_3}{\sqrt{M_{1R}}}, \eta_4=\frac{\sqrt{2} m_4}{\sqrt{M_{3R}}}$,  and $m_1, m_3$ and $m_4$ are defined in Eq.\eqref{mass matrices}. The small neutrino mass matrix for the type I+II seesaw scenario from Eq.\eqref{type II matrix} and \eqref{typeImass} is given by $\mathcal{M_\nu} = \mathcal{M^I_\nu}+\mathcal{M^{II}_\nu}$, which can be written in matrix form as 
\begin{eqnarray}
\mathcal{M_\nu}=\begin{pmatrix}
{\eta_1}^2+y_{t1} x_1+x_2 y_{t2} & 0 & \eta_1 \eta_3\\
0 &  {\eta_1}^2+y_{t1} x_1+x_2 y_{t2} & 0\\
\eta_1 \eta_3 &0 & {\eta_3}^2+{\eta_4}^2 e^{2 i \phi}+|{y_{t1}}' x_1+ {y_{t2}}' x_2| e^{i\phi_{\Delta}} \nn
\end{pmatrix}.\\
\end{eqnarray}

With the consideration of $y_{ti}',y_{ti}, \mu_{i}$ to be complex, the phase in $y_{ti}$ is fixed by the redefinition of field. The remnant phase in the triplet interaction sector being $\phi_\Delta$, which is the relative phase in ${y_{t1}}'$ and ${y_{t2}}'$. Now the above mass matrix is reduced to a simple form as
\begin{eqnarray}
\mathcal{M_\nu}=\begin{pmatrix}
r_1 & 0 & \eta_1 \eta_3\\
0 &r_1 & 0\\
\eta_1 \eta_3 & 0 & r_2 e^{i \phi_{eff}}\\
\end{pmatrix}.
\label{nmass}
\end{eqnarray}
In the above, $r_1={\eta_1}^2+x_1 y_{t1}+x_2 y_{t2}$ and $r_2=|{\eta_3}^2+{\eta_4}^2 e^{2i \phi}+|x_1 {y_{t1}}'+ x_2 {y_{t2}}'| e^{i\phi_\Delta}|$ are the $S_3$ parameters and the effective phase ($\phi_{eff}$)  in the mass matrix, which is given by  
\begin{eqnarray}
\tan{\phi_{eff}}=\frac{{\eta_4}^2 \sin{2 \phi}+|x_1 {y_{t1}}'+ x_2 {y_{t2}}'| \sin{\phi_\Delta}}{{\eta_3}^2+{\eta_4}^2 \cos{2 \phi}+|x_1 {y_{t1}}'+x_2 {y_{t2}}'| \cos{\phi_\Delta}}.
\end{eqnarray}

Since the mass matrix is already in block diagonal form, it is easy to diagonalize the only non-diagonal block by simple orthogonal rotation. The rotation matrix in $3\time 3$ dimensional form is given by 
\begin{eqnarray}
U_\nu=\begin{pmatrix}
\cos{\theta} & 0 & \sin{\theta} e^{-i \phi_\nu}\\
0  &  1 & 0\\
-\sin{\theta} e^{i\phi_\nu}&  0 & \cos{\theta}\\
\end{pmatrix}, \hspace{3mm} \text{and} \hspace{2mm}
{U_\nu}^T \mathcal{M_\nu} U_\nu = 
\begin{pmatrix}
m_{\nu_1} e^{i \phi_1} &0 &0\\
0& m_{\nu_2}  & 0\\
0 & 0 & m_{\nu_3} e^{i \phi_3}\\
\end{pmatrix}.
\label{nmassmatrix}
\end{eqnarray}
After diagonalizing the neutrino mass matrix in Eq.\eqref{nmass}, we can write the complex mass parameters of Eq.\eqref{nmassmatrix} as,
\begin{eqnarray}
&& M_{\nu_1} = m_{\nu_1} e^{i \phi_1} =\frac{r_1 +r_2 e^{i \phi_{eff}}}{2}-\frac{1}{2}\left[(r_1- r_2 e^{i \phi_{eff}})^2+4(\eta_1 \eta_3)^2\right]^{1/2}, \nn \\
&& M_{\nu_3} = m_{\nu_1} e^{i \phi_1} =\frac{r_1 +r_2 e^{i \phi_{eff}}}{2}+\frac{1}{2}\left[(r_1- r_2 e^{i \phi_{eff}})^2+4(\eta_1 \eta_3)^2\right]^{1/2}, \nn \\
&& M_{\nu_2} =r_1, \hspace{4mm} \tan{\phi_\nu}=\frac{r_2 \sin{\phi_{eff}}}{[r_1-r_2\cos{\phi_{eff}}]}.
\label{nmasscomplex} 
\end{eqnarray}
Where, $\phi_\nu$ is the only phase appearing in the neutrino mixing matrix in Eq.\eqref{nmassmatrix} and hence can be considered as Dirac like phase. $\phi_1$ and $\phi_3$ are the Majorana like phases in the mass matrix.
Hence by the standard parameterization of PMNS matrix, which is $U_{PMNS}={U_{el}}^\dagger U_\nu$, we can construct the neutrino mixing matrix for this model from Eq.\eqref{lepton mixing matrix} and Eq.\eqref{nmassmatrix} as
\begin{eqnarray}
U_{PMNS} \approx \begin{pmatrix}
\frac{x \cos{\theta}}{\sqrt{2(1-x^2)}} && \frac{1}{\sqrt{2(1+x^2)}} && \frac{ e^{-i \phi_\nu} x \sin{\theta} }{\sqrt{2(1-x^2)}}\\
-\frac{x \cos{\theta}}{\sqrt{2(1-x^2)}}-\frac{e^{i \phi_\nu} \sin{\theta} }{\sqrt{2(1+\sqrt{z})}} && -\frac{1}{\sqrt{2(1+x^2)}} && \frac{\cos{\theta}}{\sqrt{2(1+\sqrt{z})}}-\frac{ e^{-i \phi_\nu} x \sin{\theta} }{\sqrt{2(1-x^2)}}\\
\frac{\sqrt{1-2 x^2} \cos{\theta}}{\sqrt{1-x^2}}-\frac{e^{i \phi_\nu} \sqrt{z} \sin{\theta} }{\sqrt{1+\sqrt{z}}} && \frac{x}{\sqrt{1+x^2}} && \frac{\sqrt{z} \cos{\theta}}{\sqrt{1+\sqrt{z}}}+\frac{e^{- i \phi_\nu}  \sqrt{1-2 x^2} \sin{\theta} }{\sqrt{1-x^2}}
\end{pmatrix}.
\label{eq:upmns}
\end{eqnarray}
\vspace{5mm}
The mixing angles can be obtained by comparing with the standard $U_{PMNS}$ matrix as
\begin{eqnarray}
&& \sin{\theta_{13}} =|U_{e3}| \approx \frac{x \sin{\theta}}{\sqrt{2(1-x^2)}},\hspace{4mm}
\tan{\theta_{12}} =|\frac{U_{e2}}{U_{e1}}| \approx \frac{\sqrt{1-x^2}}{x \cos{\theta} \sqrt{1+x^2}}, \nn \\
\nn\\
&&\tan{\theta_{23}} =|\frac{U_{\mu 3}}{U_{\tau 3}}| \approx | \frac{\sqrt{2(1-x^2)} \cos{\theta}-\sqrt{2(1+\sqrt{z}) x\sin{\theta}} e^{-i\phi_\nu}}{\sqrt{2z(1-x^2)} \cos{\theta}+\sqrt{(1-2x^2)}\sqrt{2(1+\sqrt{z})}\sin{\theta} e^{-i\phi_\nu}}|.\\ \nn
\end{eqnarray}
For a sample value of $\theta=\frac{\pi}{6}$, we found $\sin{\theta_{13}}\approx 0.004$, $\sin^2{\theta_{12}}\approx 0.6$ and $\sin^2{\theta_{23}}\approx 0.42 $.
\subsection*{Neutrino Oscillation Parameters}
\begin{table}[htbp]
\centering
{
\renewcommand{\arraystretch}{1.2}
\catcode`?=\active \def?{\hphantom{0}}
\begin{minipage}{\linewidth}
\begin{tabular}{|l|c|c|c|}
\hline
parameter & best fit $\pm$ $1\sigma$ &  2$\sigma$ range& 3$\sigma$ range
\\
\hline\hline
$\Delta m^2_{21}\: [10^{-5}\eVq]$ & 7.56$\pm$0.19  & 7.20--7.95 & 7.05--8.14 \\
\hline
$|\Delta m^2_{31}|\: [10^{-3}\eVq]$ (NO) &  2.55$\pm$0.04 &  2.47--2.63 &  2.43--2.67\\
$|\Delta m^2_{31}|\: [10^{-3}\eVq]$ (IO)&  2.47$^{+0.04}_{-0.05}$ &  2.39--2.55 &  2.34--2.59 \\
\hline
$\sin^2\theta_{12} / 10^{-1}$ & 3.21$^{+0.18}_{-0.16}$ & 2.89--3.59 & 2.73--3.79\\
\hline
  $\sin^2\theta_{23} / 10^{-1}$ (NO)
	  &	4.30$^{+0.20}_{-0.18}$ 
	& 3.98--4.78 \& 5.60--6.17 & 3.84--6.35 \\
  $\sin^2\theta_{23} / 10^{-1}$ (IO)
	  & 5.98$^{+0.17}_{-0.15}$ 
	& 4.09--4.42 \& 5.61--6.27 & 3.89--4.88 \& 5.22--6.41 \\
\hline 
$\sin^2\theta_{13} / 10^{-2}$ (NO) & 2.155$^{+0.090}_{-0.075}$ &  1.98--2.31 & 1.89--2.39 \\
$\sin^2\theta_{13} / 10^{-2}$ (IO) & 2.155$^{+0.076}_{-0.092}$ & 1.98--2.31 & 1.90--2.39 \\
    \hline
  \end{tabular}
  \caption{ \label{tab:sum-2017} 
   The experimental values of neutrino oscillation parameters for $1\sigma$, $2\sigma$ and $3\sigma$ range \cite{deSalas:2017kay}.}
    \end{minipage}
  }
\end{table}
Considering the conventional mass ordering of three active neutrinos, one can write the cases as,\\  
\textbf{Normal Hierarchy} $\Delta {m^2_{31}} >0 $, which gives $m_1<<m_2<<m_3 $,
\begin{eqnarray*}
m_2=\sqrt{{m^2_1}+\Delta {m^2_{\rm sol}}}, \hspace{6mm} m_3=\sqrt{{m^2_1}+\Delta {m^2_{\rm atm}}}.
\end{eqnarray*}
\textbf{Inverted Hierarchy} $\Delta {m^2_{31}} <0 $, which is given by $m_3<<m_1<<m_2 $,
\begin{eqnarray*}
m_1=\sqrt{{m^2_3}+\Delta {m^2_{\rm atm}}}, \hspace{6mm} m_2=\sqrt{{m^2_3}+\Delta {m^2_{\rm atm}}+\Delta {m^2_{\rm sol}}}.
\end{eqnarray*}
\subsection{Numerical Analysis}
Now redefining the parameters appearing in the complex neutrino masses in Eq.\eqref{nmasscomplex} as $\rho_1=\frac{r_2}{r_1}$ and $\rho_2=\frac{\eta_1 \eta_3}{r_1}$ . We have the physical masses of the active neutrinos and corresponding Majorana phases in terms of new parameters as
\begin{eqnarray}
&& m_{\nu_1}=|M_{\nu_1}|= |\frac{r_1}{2}|\left[(1+\rho_1 \cos{\phi_{eff}}-C)^2+(\rho_1\sin{\phi_{eff}}-D)^2\right]^{1/2}, \nn \\
&& m_{\nu_2}=|M_{\nu_2}|= |r_1|, \nn \\
&& m_{\nu_3}=|M_{\nu_3}|=|\frac{r_1}{2}|\left[(1+\rho_1 \cos{\phi_{eff}}+C)^2+(\rho_1\sin{\phi_{eff}}+D)^2\right]^{1/2},\\
&& \tan{\phi_1}=\frac{\rho_1 \sin{\phi_{eff}}-D}{\rho_1 \cos{\phi_{eff}}-C},\nn \\
&& \tan{\phi_3}=\frac{\rho_1 \sin{\phi_{eff}}+D}{\rho_1 \cos{\phi_{eff}}+C},
\end{eqnarray}
where, 
\begin{eqnarray}
&& C=\left( \frac{A+\sqrt{A^2+B^2}}{2}\right)^{1/2}, \hspace{3mm}  D=\left( \frac{-A+\sqrt{A^2+B^2}}{2}\right)^{1/2}, \nn \\
&& A=1-2\rho_1 \cos{\phi_{eff}}+{\rho_1}^2 \cos{2 \phi_{eff}}+4{\rho_2}^2, \hspace{2mm} B=-2\rho_1 \sin{\phi_{eff}}+{\rho_1}^2 \sin{2 \phi_{eff}}.
\end{eqnarray}
And $\phi_1$ and $\phi_2$ are the Majorana like phases. We prefer a normal ordering of active neutrino masses expressed in the above equations as functions of $S_3$ parameters.
\vspace{2mm}
\begin{eqnarray}
r=\left[\frac{\Delta m^2_{\rm sol}}{\Delta m^2_{\rm atm}}\right]=\frac{1}{4} \left(2-\frac{-3+C^2+D^2+{\lambda_1}^2+2{\lambda_1} \cos{\phi_{eff}}}{C+C\lambda_1 \cos{\phi_{eff}+D \lambda_1 \sin{\phi_{eff} }}}\right)\approx 0.03.
\end{eqnarray}

\begin{figure}
\includegraphics[height=50mm,width=70mm]{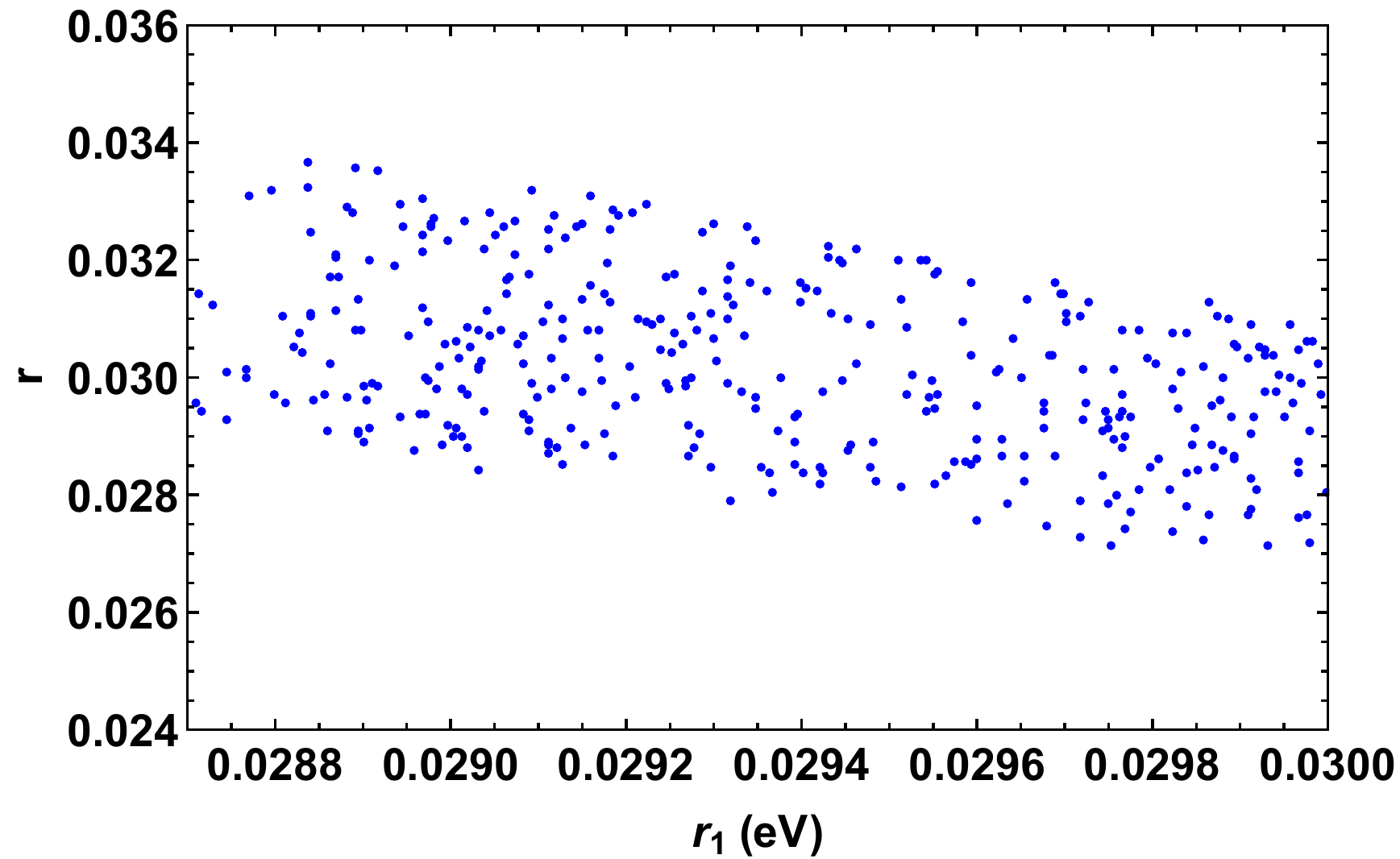}\hspace{4mm}
\includegraphics[height=50mm,width=70mm]{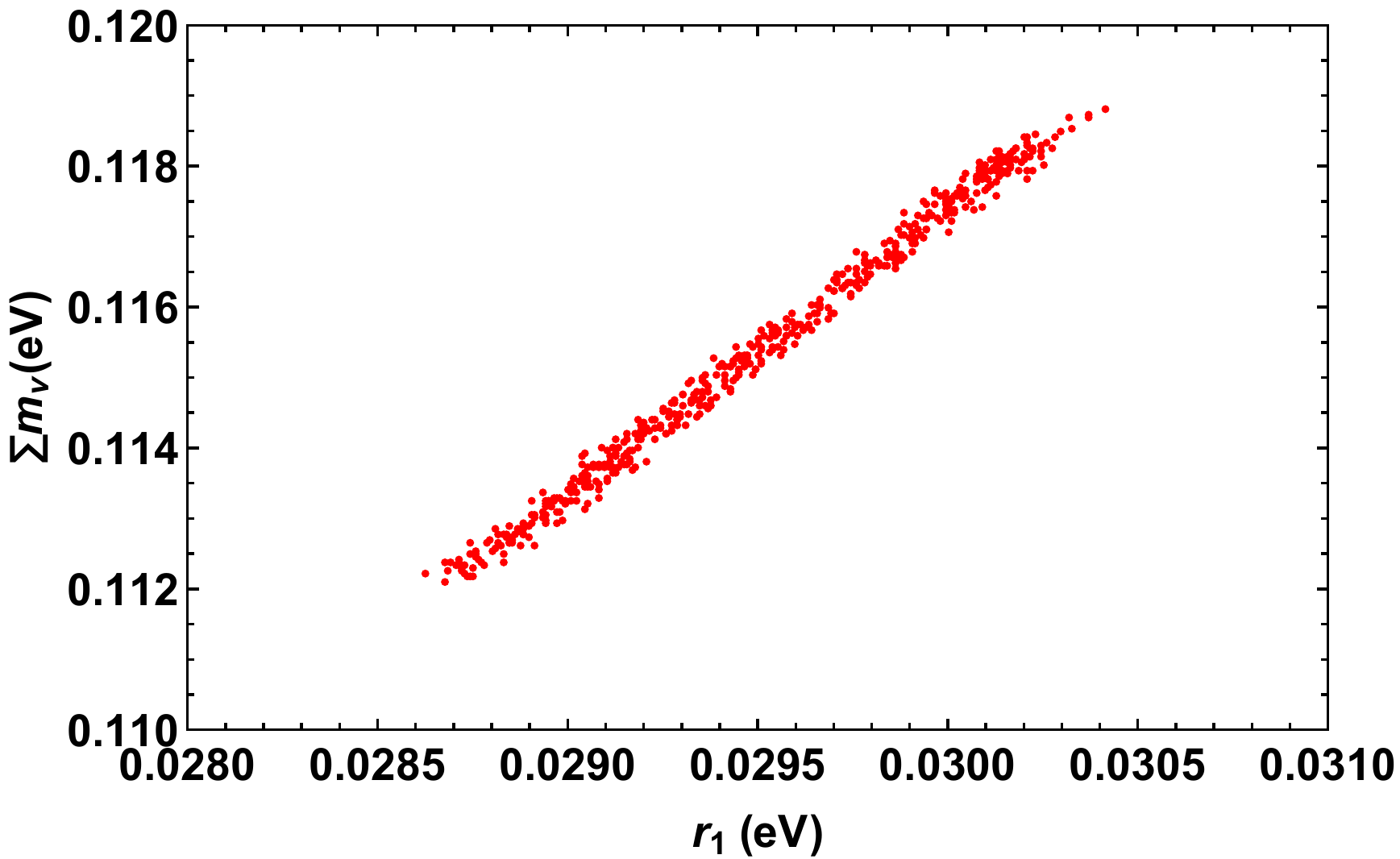}
\caption{The left panel displays the variation of ratio of solar to atmospheric mass squared differences with $r_1$ and the right panel shows the variation of total active neutrino mass with the same.}
\label{r1}
\end{figure}

\begin{figure}
\includegraphics[height=50mm,width=70mm]{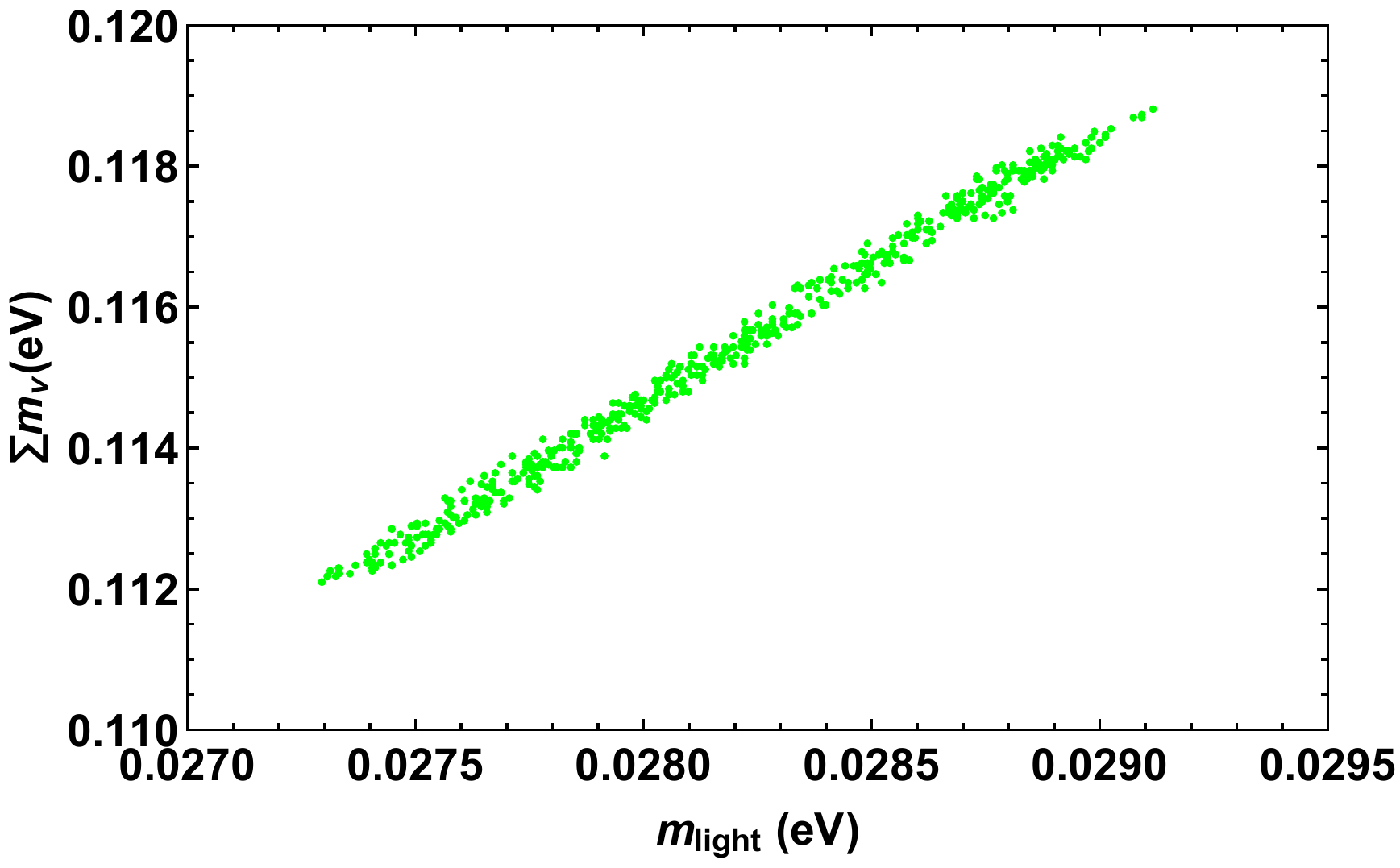}\hspace{4mm}
\includegraphics[height=50mm,width=70mm]{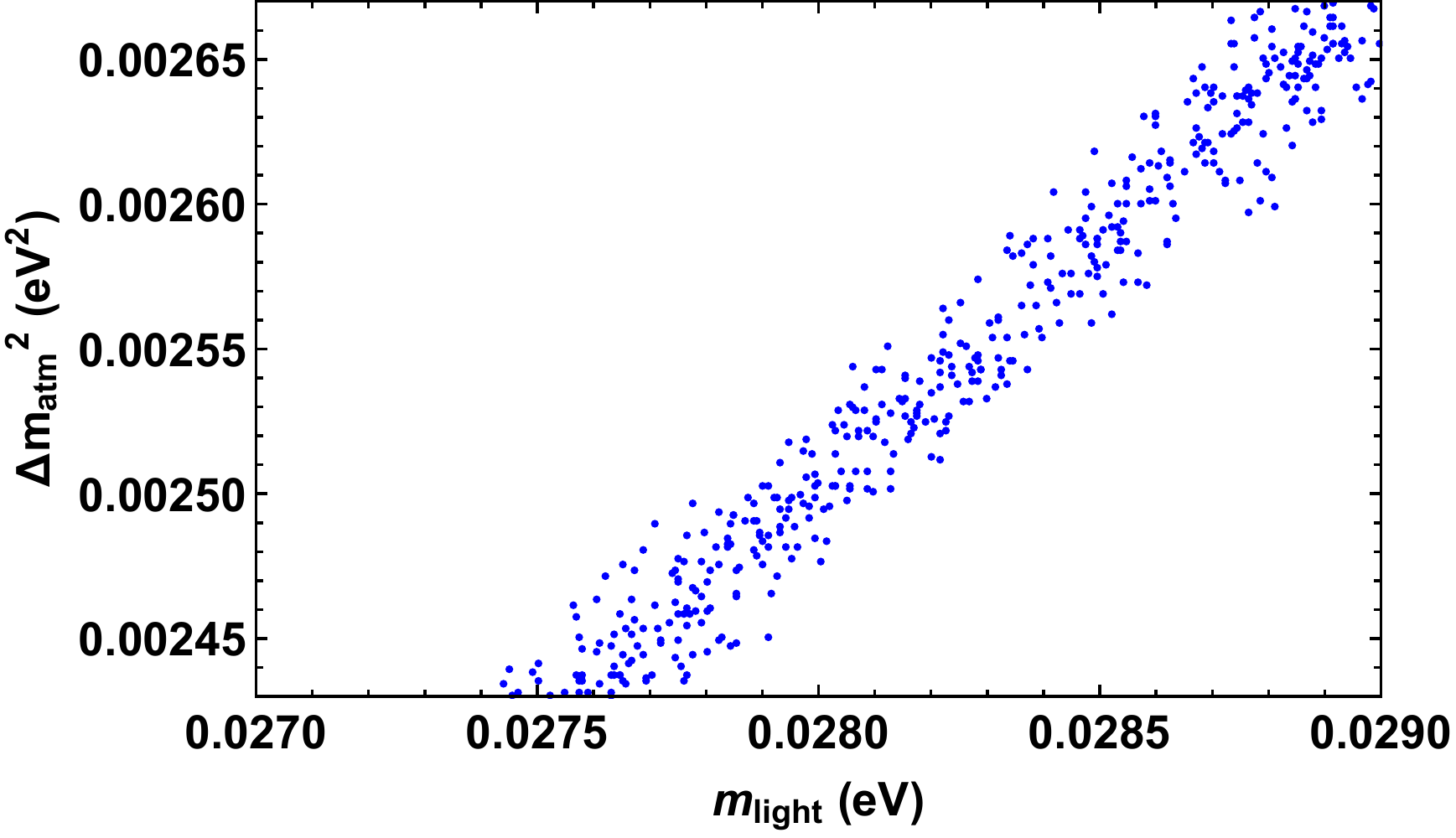}
\caption{Variation of the total active neutrino mass (left panel) and atmospheric mass squared difference (right panel) with the lightest neutrino mass in the model.}
\label{mlight}
\end{figure}

\begin{figure}
\includegraphics[height=50mm,width=70mm]{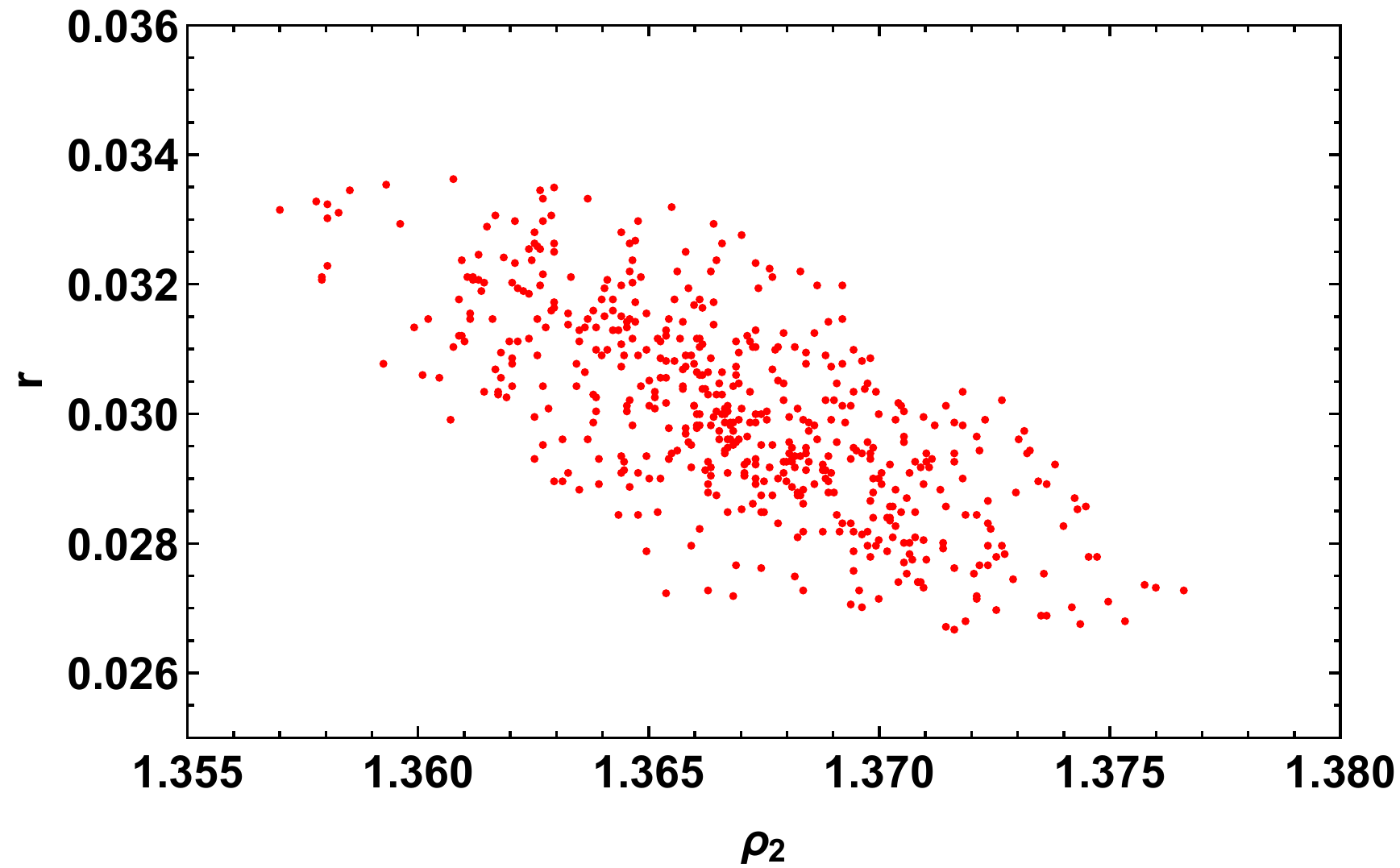}\hspace{4mm}
\includegraphics[height=50mm,width=70mm]{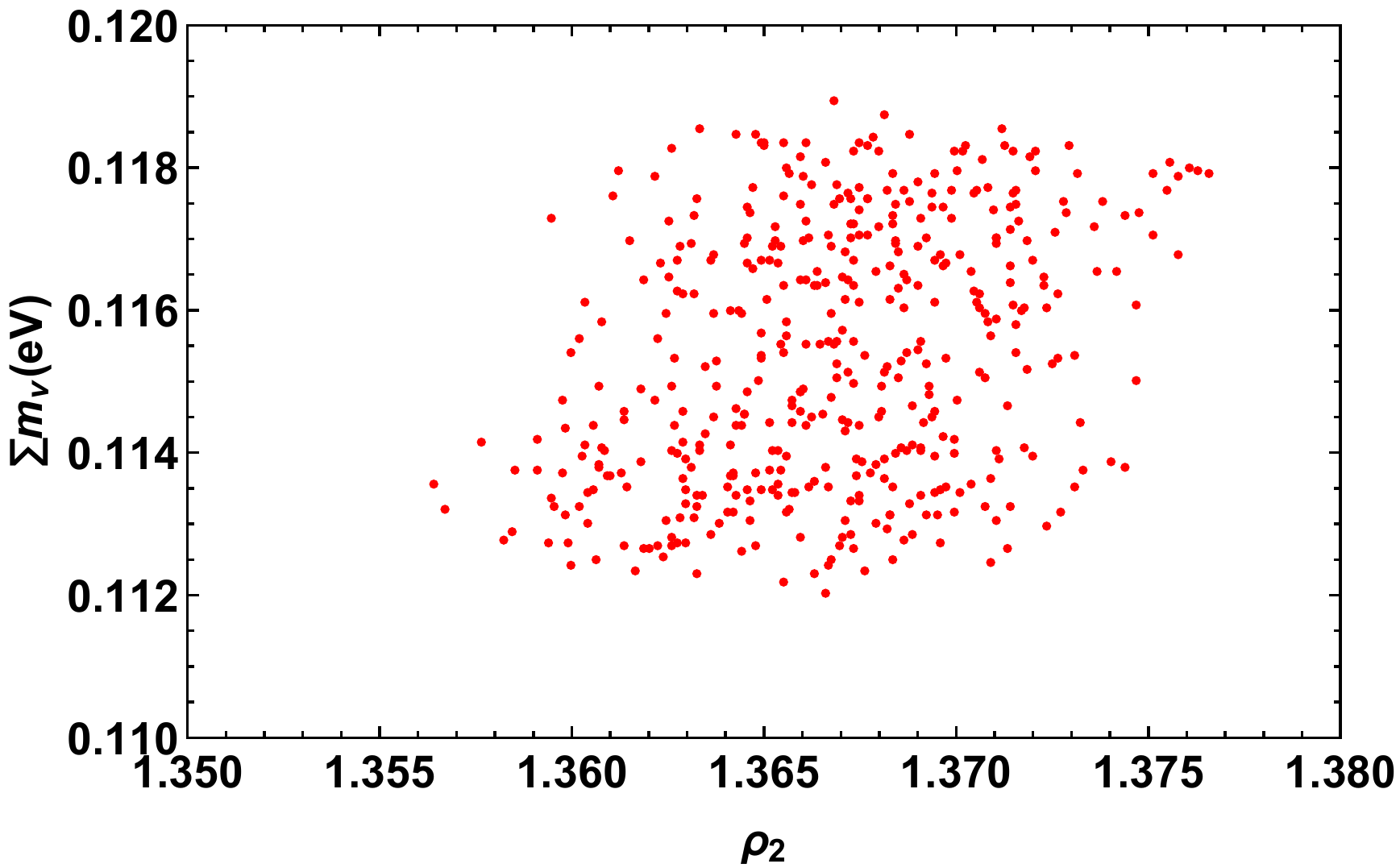}
\caption{Variation of ratio of solar to atmospheric mass squared difference (left panel) and sum of active neutrino masses (right panel) with $\rho_2$.}
\label{rho2}
\vspace{4mm}
\includegraphics[height=50mm,width=70mm]{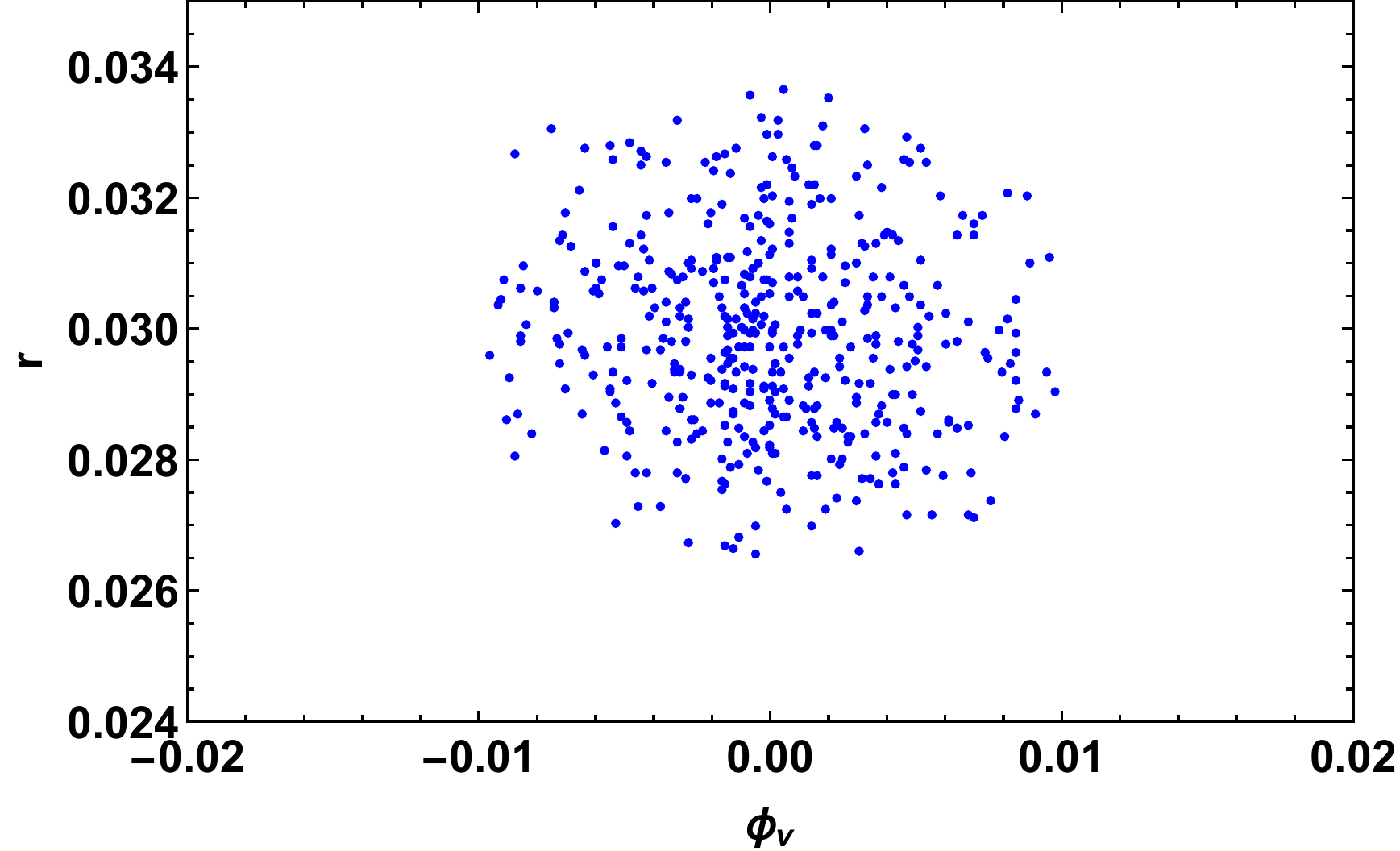}\hspace{4mm}
\includegraphics[height=50mm,width=70mm]{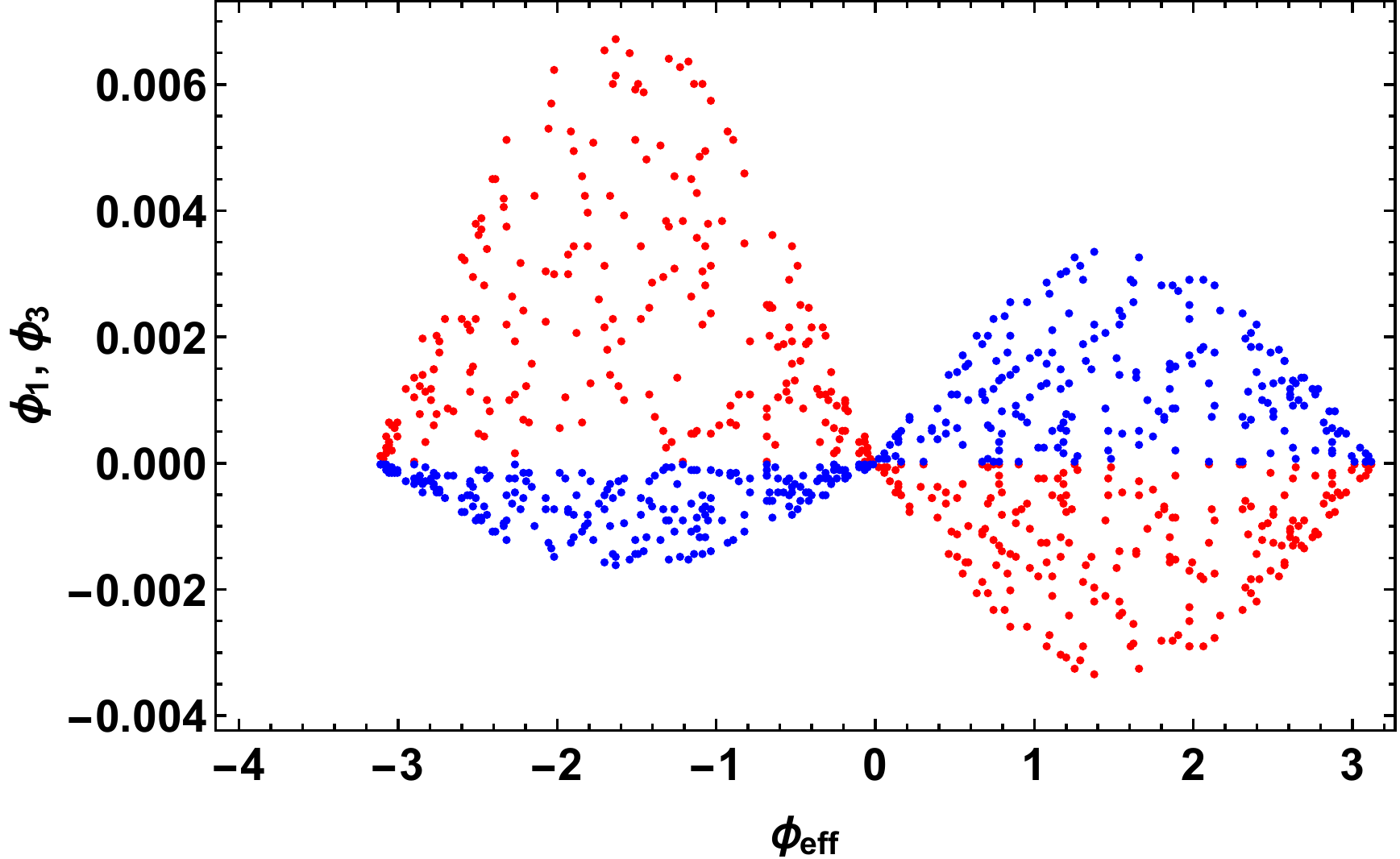}
\caption{Left panel represents the parameter space for Dirac like phase $\phi_\nu$, allowed by the observed solar to atmospheric neutrino mass ratio and the right panel shows the correlation of remnant phase in the mass matrix with the Majorana like phases.}
\label{phinu}
\end{figure}

\hspace{5mm}We vary the model parameters $r_1$ from 0 to $0.1$, $\rho_1$ from 0 to $0.01$ and $\rho_2$ from 0 to 2 and shown the allowed parameter space for those parameters compatible with the current $3\sigma$ data of neutrino mass squared differences and cosmological bound on total active neutrino masses. The left and right panel of Fig.\ref{r1} represent the variation of $r_1$ with the ratio of solar to atmospheric mass squared differences and the total neutrino mass, respectively, which in turn gives a strong constraint on $r_1$ to lie within a range $0.027$ to $0.03$. Similarly, the lightest neutrino mass in the model is constrained to vary from $0.025$ to $0.029$ as shown in Fig.\ref{mlight}. The topmost panel of the Fig.\ref{rho2}, shows the variation of $\rho_2$ with the ratio of solar to  atmospheric mass squared differences, which constrain the parameter space of $\rho_2$ to lie within the range $1.35$ to $1.38$. We also show the variation of Dirac like phase ($\phi_\nu$), appearing in the neutrino mixing matrix ($U_{PMNS}$) in Eq.\eqref{eq:upmns} with $ r$, which gives a parameter space allowed to lie within $-0.01$ to $0.01$, and is very small.

\section{Leptogenesis}
Leptogenesis, through the out of equilibrium decay of heavy particles, is the most viable way to generate observed baryon asymmetry of the universe. For a review of leptogenesis, one can refer \cite{lepreview:2005eh}. Even though this formalism is widely studied with the heavy Majorana fermions, there exist few studies, which focus on the production of asymmetry through scalar decays within the type II seesaw framework \cite{Hambye:2005tk}\cite{Sierra:2014tqa}. Here, as emphasized before, the scalar triplets are not enough to explain the observed neutrino mixing within only type II seesaw scenario. Therefore, we add extra right handed neutrinos to include type I seesaw framework for the explanation of neutrino phenomenology. Unlike the Majorana fermions, triplets are charged and can generate an asymmetry in particle-antiparticle decay width. Both scalar triplets and right-handed neutrinos in the scenario under consideration can contribute to the leptonic CP violation. Various cases are possible, which include the generation of asymmetry solely from the decay of scalar triplets and right handed neutrinos or from both, depending on their masses and are well studied in literature \cite{AristizabalSierra:2012pv}. In this work, we aim to explore the leptogenesis phenomenon from the lightest heavy triplet in a scenario of high and low energy scales. It is  pointed out earlier in the lierature that one cannot generate a non-zero CP asymmetry with one loop contribution in the presence of a single scalar triplet \cite{Felipe:2013kk}. However, adding atleast one additional triplet scalar can help resolve the problem. Before focusing on the one loop level decay of scalar triplets, we discuss the possible tree level decay channels, as the CP asymmetry is calculated from the interference of tree and the one loop level diagrams. 
\begin{figure}
\includegraphics[height=20mm,width=40mm]{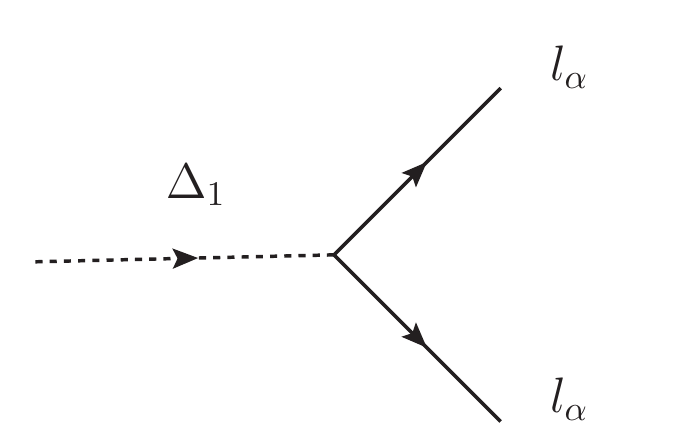}
\includegraphics[height=20mm,width=40mm]{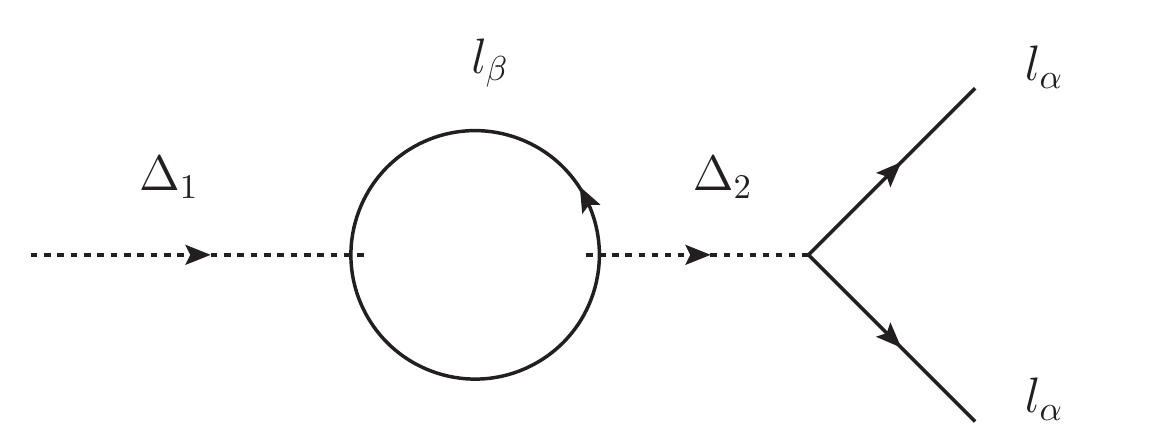}
\includegraphics[height=20mm,width=40mm]{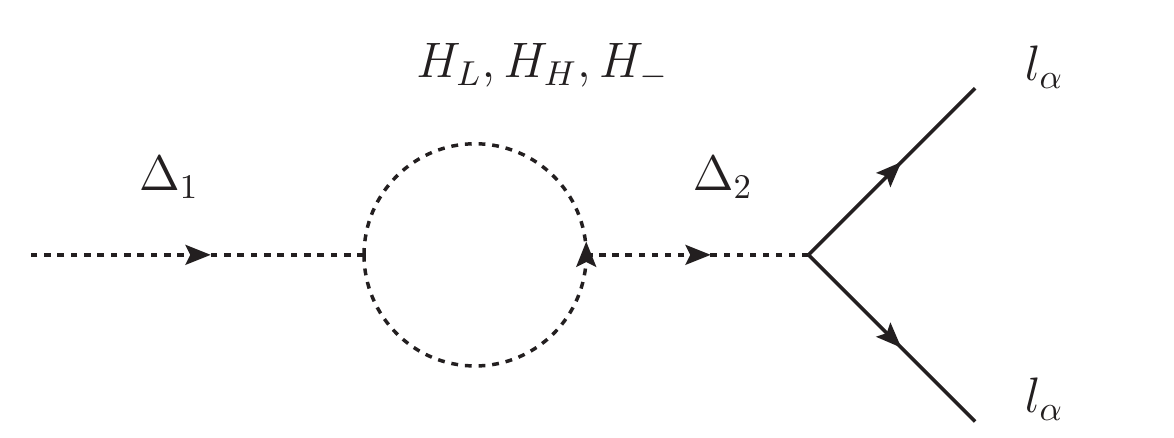}
\includegraphics[height=20mm,width=40mm]{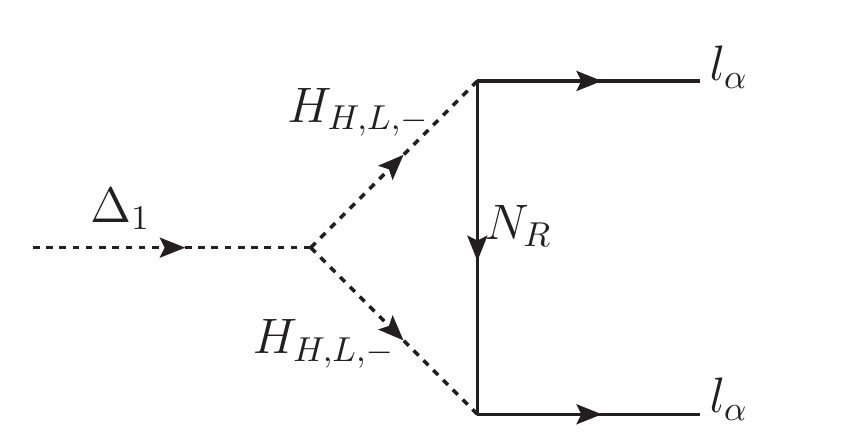}
\caption{The scalar triplet tree and 1-loop diagrams in the presence of additional triple scalars contributing to the leptogenesis} 
\label{feyntrip} 
\end{figure}
From the type II seesaw Lagrangian Eq.\eqref{Model interaction lagrangian}, one can find two possible decay modes of the scalar triplets, namely, the triplet can decay to two leptons or two scalars in the final state, which are given by ${\Delta}\rightarrow \bar{\ell_i}\bar{ \ell_i}$ and $\Delta \rightarrow h_j h_j$, where, $h_j$ are the Higgs fields of the model.\\
Summing over all the final states for leptons and Higgs decay modes of scalar triplet, we have
\begin{eqnarray}
&&\Gamma(\Delta^{--}_i \rightarrow \ell^- \ell^-)=\Gamma({\Delta^0}^\star_i \rightarrow \nu \nu)=
\Gamma(\Delta^{-}_i \rightarrow l^- \nu)= \frac{M_{\Delta i}}{8 \pi} {\rm Tr}[{{Y_{\Delta_i}}}{Y^\dagger_{\Delta_i}}].\\
&&\Gamma(\Delta^{--}_i\rightarrow h^- h^-)=\Gamma({\Delta^0}^\star_i \rightarrow {h_0}^\star {h_0}^\star)=
\Gamma(\Delta^-_i \rightarrow h^- {h_0}^\star)= \frac{1}{8 \pi {M_{\Delta_i}}}|\mu_{\Delta_i}|^2.
\label{decayoftriplet}
\end{eqnarray}
\begin{figure}
\end{figure}
Hence the branching ratios for the decay of $\Delta_1$ are given by
\begin{eqnarray}
&& B_l=\sum_{l=e,\mu,\tau} B_l=\frac{M_{\Delta_1}}{8\pi \Gamma_{tot}} Tr(Y_{\Delta_1}Y^\dagger_{\Delta_1}),\\
&& B_h=\sum_{i=L,H,-} B_{H_i}=\frac{|\mu_{1L}|^2+|\mu_{1H}|^2+|\mu_{1-}|^2}{8 \pi M_{\Delta_1} \Gamma_{\Delta_1}}=\frac{|\lambda|^2 M_{\Delta_1}}{8 \pi \Gamma_{\Delta_1}},\\
&& \Gamma_{\Delta_1}=\frac{M_{\Delta_1}}{8 \pi} \left( \sum |Y_{\Delta_1}|^2_{ii}+|\lambda|^2 \right),
\end{eqnarray}
where, $\lambda=\frac{(|\mu_{1L}|^2+|\mu_{1H}|^2+|\mu_{1-}|^2)^{1/2}}{M_{\Delta_1}}$, and $\Gamma_{\Delta_1}$ is the total decay rate of the first triplet. $\mu_{1L}$, $\mu_{1H}$ and $\mu_{1-}$ are the coupling of the decaying triplet to $H_L$, $H_H$ and $H_-$ respectively, which are defined as follows
\begin{eqnarray}
&& \mu_{1L}=\mu_1 {\sin}^2{\beta}+{\mu_1}'{\cos}^2{\beta}, \nonumber \\
&& \mu_{1H}=\mu_1 {\cos}^2{\beta}+{\mu_1}'{\sin}^2{\beta}, \nonumber \\
&& \mu_{1-}=\mu_1.
\end{eqnarray}

From the type II Lagrangian mentioned in Eq.\eqref{Model interaction lagrangian}, the scalar triplet-lepton Yukawa couplings can be written in the matrix form as
\begin{eqnarray}
Y_{\Delta_1}=\begin{pmatrix}
y_{t1} && 0 &&0\\
0 && y_{t1} &&0\\
0 && 0 && {y_{t1}}'
\end{pmatrix}, \hspace{6mm} Y_{\Delta_2}=\begin{pmatrix}
y_{t2} && 0 &&0\\
0 && y_{t2} &&0\\
0 && 0 && {y_{t2}}'
\end{pmatrix}.
\label{yukawa-triplet}
\end{eqnarray}
Since the SM leptons and right-handed neutrinos couple to different Higgs doublets of the present model, we can re-write the Yukawa coupling matrix corresponding to the interaction of these fermions with $H_L$, $H_H$ and $H_-$ from Eq.\eqref{Model interaction lagrangian} after redefinition and rotation of Higgs fields in Eq\eqref{redefhiggs} and \eqref{higgsmixing}. One can write
\begin{eqnarray}
&& Y^\nu_L=\begin{pmatrix}
\frac{y_{\nu_2}}{\sqrt{2}}\sin{\beta} && \frac{y_{\nu_2}}{\sqrt{2}}\sin{\beta} && 0\\
\frac{y_{\nu_2}}{\sqrt{2}}\sin{\beta} && -\frac{y_{\nu_2}}{\sqrt{2}}\sin{\beta} && 0\\ 
\frac{y_{\nu_3 e^{i p_{y3}}}}{\sqrt{2}}\sin{\beta} && \frac{y_{\nu_3} e^{i p_{y3}}}{\sqrt{2}}\sin{\beta} && y_{\nu_4} e^{i p_{y4}} \cos{\beta}
\end{pmatrix},\, \, Y^\nu_{-}=\begin{pmatrix}
-\frac{y_{\nu_2}}{\sqrt{2}} && \frac{y_{\nu_2}}{\sqrt{2}} && 0\\
\frac{y_{\nu_2}}{\sqrt{2}} && \frac{y_{\nu_2}}{\sqrt{2}} && 0\\
\frac{y_{\nu_3}}{\sqrt{2}} && -\frac{y_{\nu_3}}{\sqrt{2}} && 0
\end{pmatrix}, \nn
\end{eqnarray}
\begin{align}
 Y^\nu_H=\begin{pmatrix} 
\frac{y_{\nu_2}}{\sqrt{2}}\cos{\beta} & \frac{y_{\nu_2}}{\sqrt{2}}\cos{\beta} & 0\\
\frac{y_{\nu_2}}{\sqrt{2}}\cos{\beta} & -\frac{y_{\nu_2}}{\sqrt{2}}\cos{\beta} & 0\\
\frac{y_{\nu_3} e^{i p_{y3}}}{\sqrt{2}}\cos{\beta} & \frac{y_{\nu_3} e^{i p_{y3}} }{\sqrt{2}}\cos{\beta} & e^{i p_{y4}} y_{\nu_4} \sin{\beta}
\end{pmatrix}.
\label{yukawaN}
\end{align}
\\
Here we consider the leptogenesis in the mass basis of charged leptons. Hence the Dirac Yukawa matrices, mentioned above, will be modified as $\tilde{Y}^\nu_{H,L,-}=Y^\nu_{H,L,-}  U_{el}$. If we use the observed masses of the charged leptons, we can have the numerical entries of the matrix $U_{el}$ in Eq.\eqref{lepton mixing matrix}, from which  the modified Yukawa coupling for $H_L$ as\\
\begin{equation}
\tilde{Y^\nu_L}=\begin{pmatrix}
0 && 0 && y_{\nu_1} \sin{\beta}\\
0.005 y_{\nu_1}\sin{\beta} &&  y_{\nu_1} \sin{\beta} && 0\\
y_{\nu_4} \cos{\beta} e^{ip_{y_4}} && 0.005 y_{\nu_1} \sin{\beta} && 0.00001 y_{\nu_4}\cos{\beta} e^{ip_{y_4}}+y_{\nu_3}\sin{\beta} e^{ip_{y_3}}
\end{pmatrix}.
\end{equation}
\\
Similarly, we can have the modified Yukawa coupling matrices $\tilde{Y^\nu_H}$ and $\tilde{Y^\nu_{-}}$, corresponding to the Yukawa couplings of $H_H$ and $H_-$. Further the entries of modified Yukawa couplings will be denoted as $\tilde{y_{\nu_i}}$. We explore the importance of the above mentioned couplings in explaining leptogenesis in both high and low scale regimes in the following subsections. 

\subsection{Case I: Leptogenesis with $M_{\Delta_1}=\mathcal{O}(10^{10})$ GeV} 
Here we consider leptogenesis solely from scalar triplets by assuming the right-handed neutrinos to be much heavier in mass and decoupled earlier \cite{AristizabalSierra:2012pv}. At high temperature regime, the scalar triplets are thermalized because of the gauge interactions. They decouple, when the temperature of thermal bath approaches the mass of decaying triplet, i.e,  $T\approx {M_{\Delta_1}}$ and produce the lepton asymmetry. We choose the mass of the scalar triplet to be of the order of $\mathcal{O}(10^{10})$  GeV and the mass of the lightest right-handed neutrino to be of the order of $\mathcal{O}(10^{11})$ GeV. Generation of nonzero CP asymmetry from the scalar triplet decay, in one loop level, requires at least one more heavy triplet ($M_{\Delta_2} >> M_{\Delta_1}$), as mentioned in the literature. As the scalar triplet has two decay modes (to different Higgs in the model and the leptons), any of the decay channel being out of  equilibrium can generate lepton asymmetry. \\ 
\textbf{Constraints on couplings from out of equilibrium decay and neutrino mass:}\\
To generate nonzero asymmetry as per the Sakharov's condition the decay rate should be less than the Hubble expansion
\begin{eqnarray}
\Gamma_{\Delta_1} & < & H \approx 1.66 \times \sqrt{g_\star}\frac{T^2}{1.2\times 10^{19}}~\rm GeV \nn\\
&& < 1.38\times 10^{-18} [T^2]_{T=M_\Delta} (g_\star \approx 100)~\rm GeV.
\end{eqnarray}
The cosmological bound on sum of the neutrino masses is found to be less than $0.12$ eV \cite{Aghanim:2018eyx} . In this model, we are considering the total contribution to the neutrino mass contributing equally from type I and II sectors. Hence, we can constrain the Yukawa couplings with the assumption that the lightest right-handed neutrino is $\mathcal{O}(1)$ lighter in mass than the other two heavy neutrinos,\\
From type I:
\begin{eqnarray}
&& \sum m^I_\nu=\frac{4 y^2_{\nu_1} v^2 \sin^2{\beta}}{M_{1R}}+\frac{2 y^2_{\nu_3} v^2 \sin^2{\beta}}{M_{1R}}+\frac{2 y^2_{\nu_4} v^2 \cos^2{\beta}}{M_{1R}}\leq 0.05\times 10^{-9}~\rm GeV \nn\\
&& \approx 0.3 y^2_{\nu_3}\sin^2{\beta}+y^2_{\nu_4} \cos^2{\beta} \leq \frac{(0.05\times 10^{-9}) M_{3R}}{2 v^2}.
\end{eqnarray}
From type II:
\begin{eqnarray}
&& \sum_i m^{II}_{\nu i}=\frac{(4y_{ti}+2{y_{ti}}')\mu_{iL} v^2}{M^2_{\Delta_i}} \leq 0.05\times 10^{-9}~ \rm GeV \nn \\
&& \approx (2 y_{ti}+{y_{ti}}')\mu_{iL} \leq \frac{(0.05\times 10^{-9} ) M^2_{\Delta_i}}{2 v^2} ~\rm GeV.
\end{eqnarray}
The general expression for CP asymmetry from the leptonic self energy and vertex contribution to the scalar triplet decay is provided below
\begin{eqnarray}
{\epsilon^{\ell}_\Delta}=\sum_i \frac{M^2_{\Delta_1}}{2 \pi} \frac{{\rm Im}\left[(Y^\dagger_{\Delta_1} Y_{\Delta_2})_{ii} {\rm Tr}(Y^\dagger_{\Delta_1} Y_{\Delta_2})\right] }{M^2_{\Delta_1} {\rm Tr}[{Y_{\Delta_1}} Y^\dagger_{\Delta_1}]+ \mu^2_{\Delta_1}} g\left(\frac{M^2_{\Delta_1}}{M^2_{\Delta_2}}\right),
\end{eqnarray}
where, \begin{equation}
g(x)=\frac{x(1-x)}{(1-x)^2+xy},\hspace{3mm} x=\frac{M^2_{\Delta_1}}{M^2_{\Delta_2}} \hspace{2mm} \text{and} \hspace{2mm} y=\left(\frac{\Gamma_{\Delta_2}}{M_{\Delta_2}}\right)^2.
\end{equation}
\begin{figure}
\includegraphics[height=50mm,width=70mm]{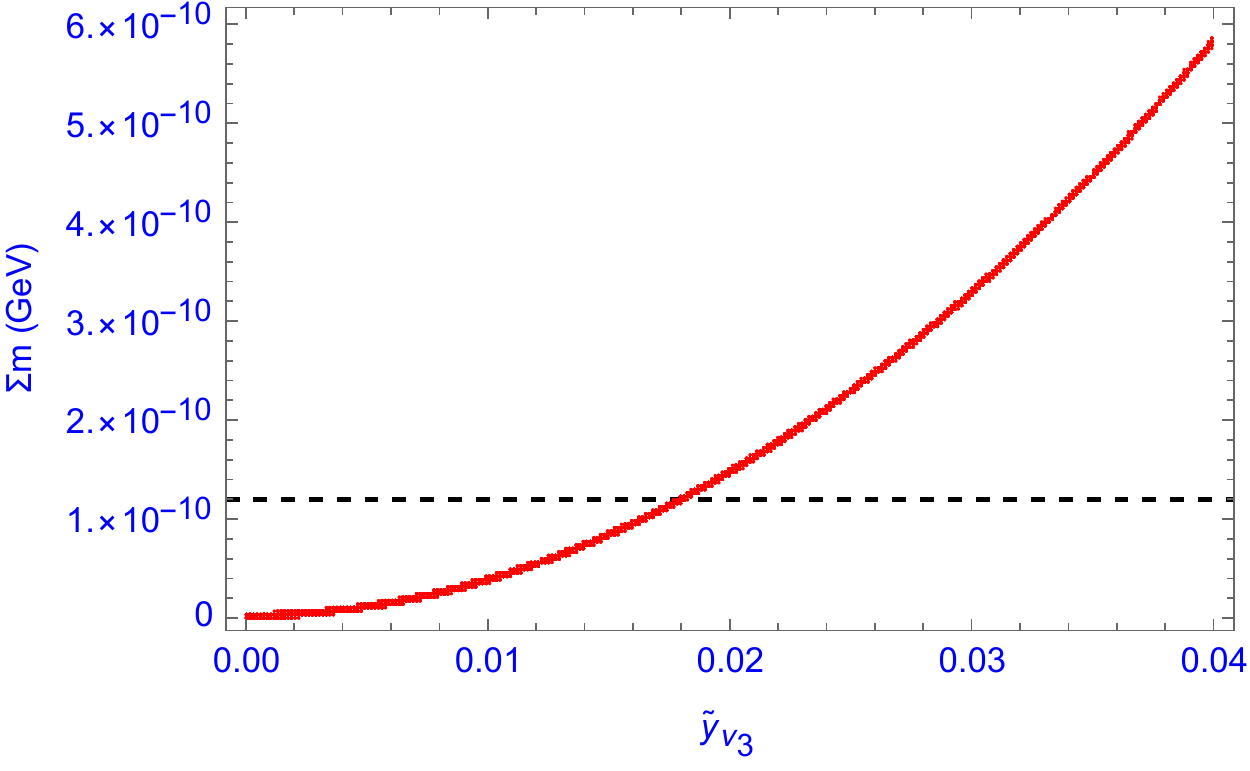}\hspace{4mm}
\includegraphics[height=50mm,width=70mm]{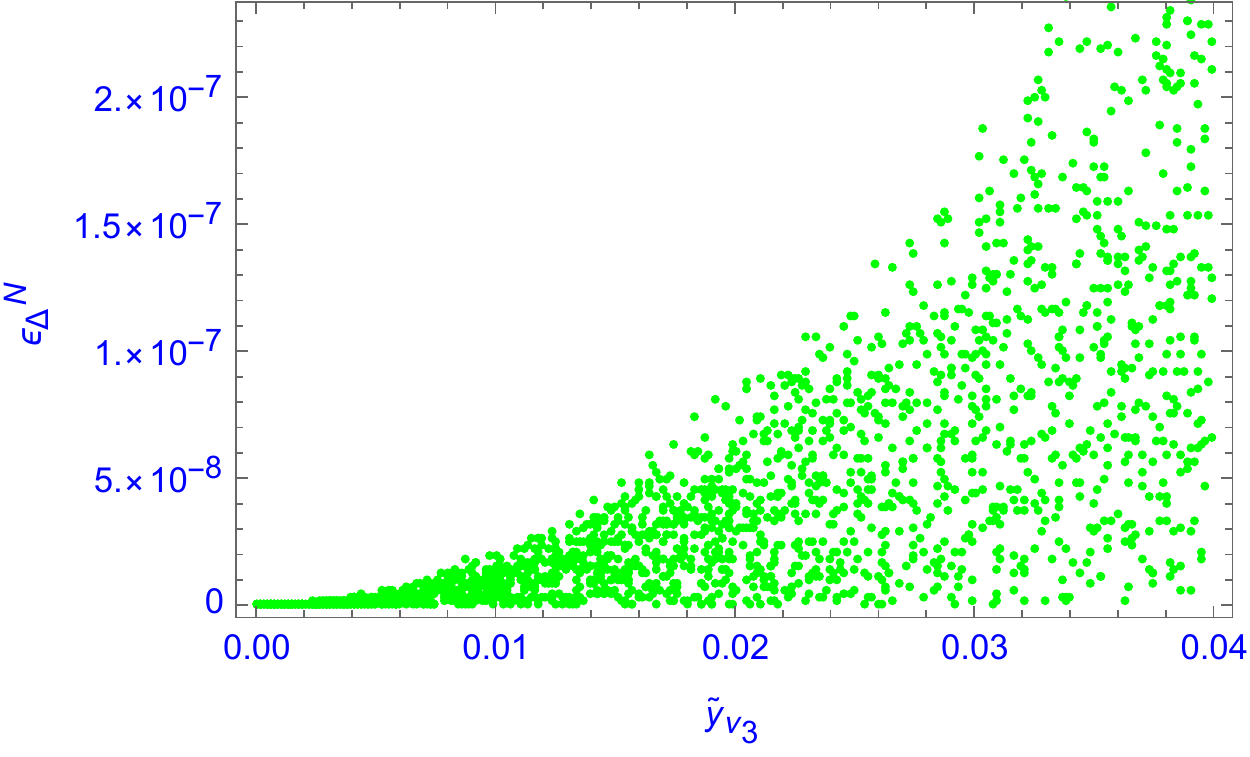}
\caption{Left panel shows the variation of Yukawa coupling with the sum of neutrino masses in compatible with the 3$\sigma$ neutrino oscillation parameters and cosmological bound on total neutrino mass. Right panel shows the variation of same Yukawa coupling with the CP asymmetry generated, satisfying the observed baryon asymmetry of the universe.}
\label{cpn}
\vspace{6mm}
\includegraphics[height=50mm,width=70mm]{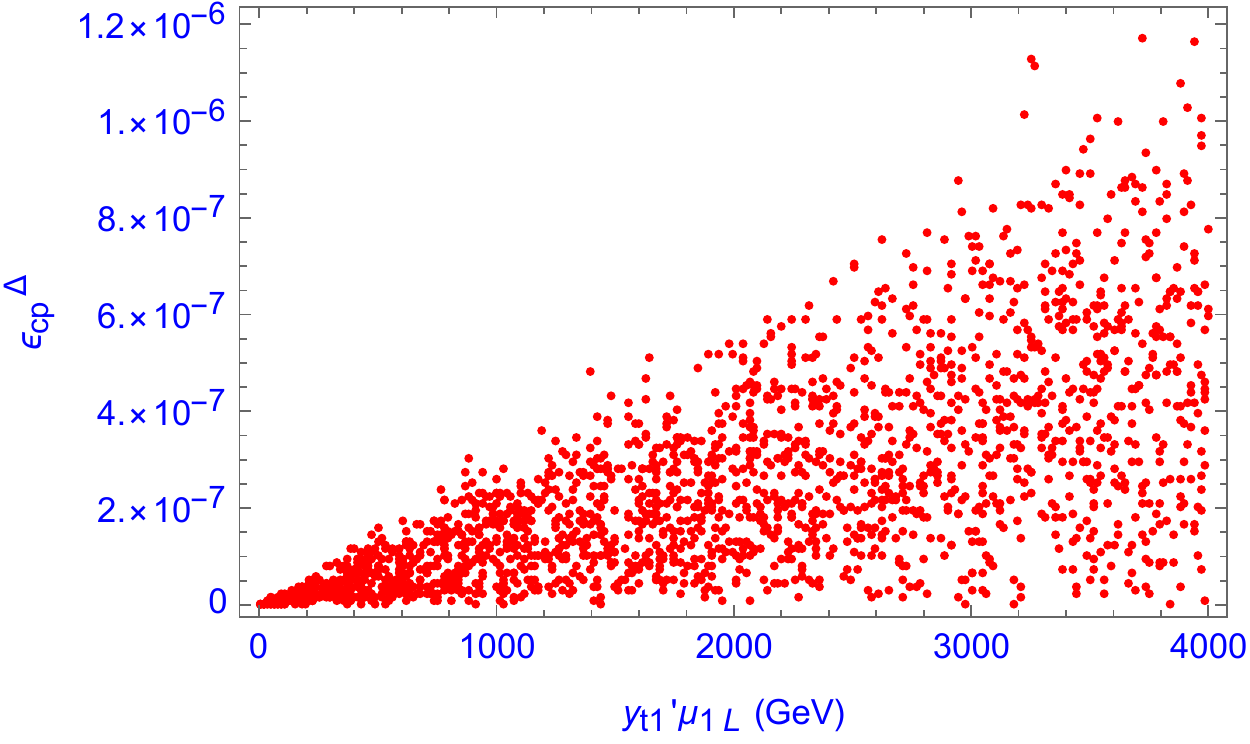}\hspace{4mm}
\includegraphics[height=50mm,width=70mm]{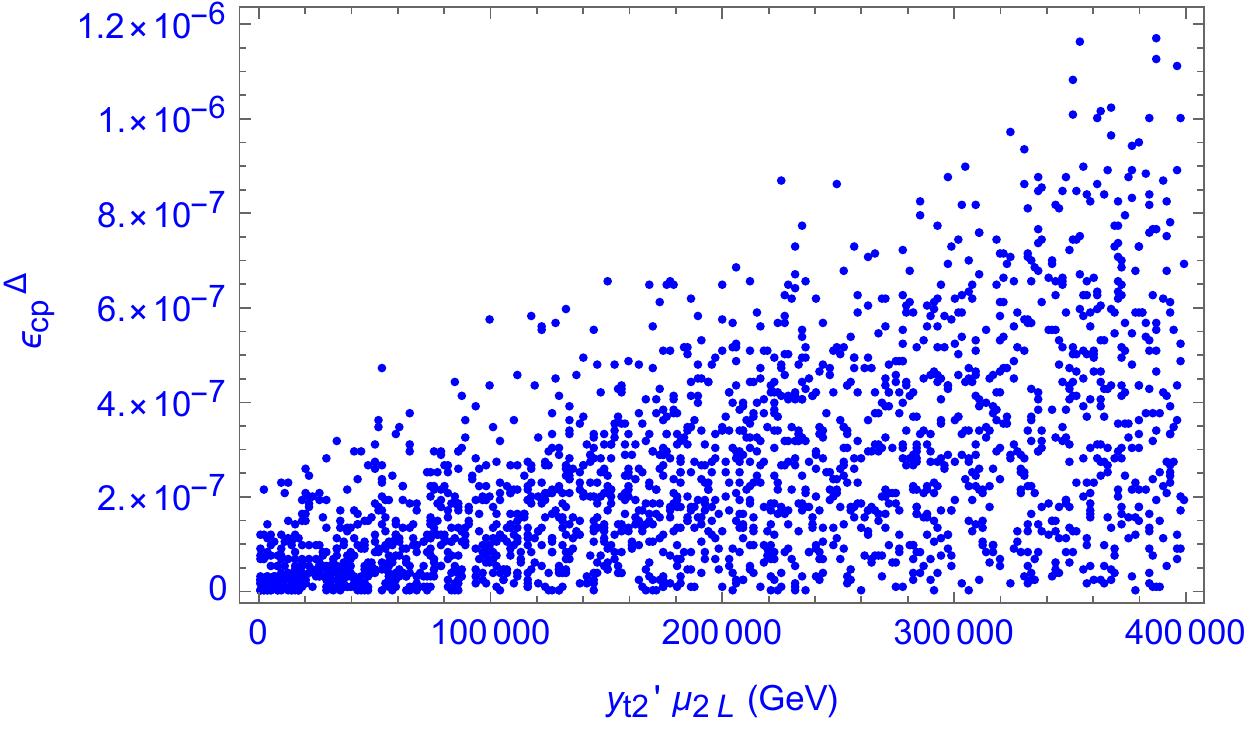}
\caption{Left panel displays the variation of total CP asymmetry, generated by the first triplet with the corresponding triplet-lepton Yukawa coupling and right panel represents the variation of CP asymmetry with the Yukawa coupling for second triplet, compatible with the observed Baryon asymmetry and neutrino mass. }
\label{cph}
\end{figure}
Hence the total CP asymmetry from the lepton loop in one flavor regime is  $\epsilon_l=\sum_{\alpha=e,\mu,\tau}\epsilon^\alpha_l=0$, due to the diagonal triplet-lepton Yukawa matrix. This structure of Yukawa also disfavors the flavored leptogenesis scenario. Hence the only self energy diagram that contributes to the triplet leptogenesis scenario is through Higgs loop, as shown in Fig.\ref{feyntrip}. Here the scalar triplet to Higgs decay mode to be out of equilibrium as per the Sakharov's conditions. This leads to $B_h \Gamma_{tot}<H$, where H is the Hubble expansion rate of the universe. CP asymmetry from the interference of tree and scalar self energy loop is given by
\begin{eqnarray}
\epsilon^h_\Delta &=&\sum_{i} \frac{1}{2 \pi} \frac{{\rm Im} (Y^\dagger_{\Delta_1} Y_{\Delta_2})_{ii} {\mu^\star_{\Delta_1}} \mu_{\Delta_2}}{M^2_{\Delta_1} Tr[Y^\dagger_{\Delta_1} Y_{\Delta_1}]+ {|\mu_{\Delta_1}|}^2}  g\left(\frac{M^2_{\Delta_1}}{M^2_{\Delta_2}}\right)\nn \\
&& \approx \frac{1}{16 \pi^2}\frac{{\rm Im} [{y_{t1}}'{y_{t2}}'(\mu_{1L}\mu^\star_{2L}+\mu_{1H}{\mu^\star_{2H}}+\mu_{1-}{\mu^\star_{2-}})]}{\Gamma_{\Delta_1} M_{\Delta_1}}   g\left(\frac{M^2_{\Delta_1}}{M^2_{\Delta_2}}\right) \nn\\
&& \approx \frac{3}{16 \pi^2}\frac{3{y_{t1}}'{y_{t2}}'(\mu_{1L}{\mu^\star_{2L}}) \sin{\phi_{\rm cph}}}{\Gamma_{\Delta_1} M_{\Delta_1}} g\left(\frac{M^2_{\Delta_1}}{M^2_{\Delta_2}}\right).
\label{CPH}
\end{eqnarray}
Since there is no contribution to CP asymmetry from the lepton loop in one flavor approximation, one can generate a large CP asymmetry from scalar self energy loop and vertex diagrams with right handed neutrinos. CP asymmetry from the right handed neutrino loop is given by
\begin{eqnarray}
\epsilon^N_{\Delta}&=&\frac{-1}{4\pi}\sum_i M_{iR} \frac{Im\left[\mu_{\Delta_1} ({Y_{\Delta_1}})_{ii}(\tilde{Y}^{\nu \star}_{L,H,-} \tilde{Y}^{\nu \star}_{L,H,-})_{ii} \right]}{M^2_{\Delta_1} Tr[Y_{\Delta_1}Y^\dagger_{\Delta_1}]+\mu^2_{\Delta_1}} {\rm ln}\left(1+\frac{M^2_{\Delta_1}}{M^2_{iR}}\right) \nn \\
&&\approx\frac{-1}{4\pi}\frac{{M^2_{\Delta_1}}}{M_{3R}} \frac{\mu_{\Delta_1}{y_{t1}}'\tilde{y_{\nu_3}}^2\sin{\phi_{\rm cpn}}}{{M_{\Delta_1}}^2 Tr({Y_{\Delta_1}}^\dagger Y_{\Delta_1})+|\mu_{\Delta_1}|^2}\hspace{3mm}(M_{iR}>>M_{\Delta_1}).
\label{CPN}
\end{eqnarray}
\begin{table}[h]
\begin{center}
\begin{tabular}{| c | c | c | c | c | c| c| c|}
\hline
~Parameters~ & $M_{\Delta_1}$(GeV) & $M_{N}$(GeV) & $\tilde{y_{\nu_3}}$ & ${y_{t1}}'{\mu_L}$ &~ $\sum{m_\nu}$(eV) ~& ~$\epsilon_{CP} =\epsilon^N_{\Delta}+\epsilon^h_\Delta$ \\
\hline
 BP1 & $10^{10}$ & $2.1\times 10^{11}$ & $0.02$ & $1.9\times 10^3$  & $0.07$ & $7.6 \times 10^{-7}$\\
 \hline
 BP2 & $10^{10}$& $2.1\times10^{11}$ & $0.03$ & $6.4\times 10^3$ & $ 0.09$ & $5.8\times10^{-7}$\\
   \hline
\end{tabular}
\end{center}
\caption{Some sample benchmark points (BP) for the couplings are provided by using Eq.\eqref{CPN} and \eqref{CPH}, which satisfy both neutrino mass and leptogenesis simultaneously. }
\label{bp1}
\end{table}
Here, we consider the constraint on the relevant Dirac Yukawa coupling ($y_{\nu_3}$) from type I seesaw neutrino mass, by fixing the Higgs mixing angle $\beta=\frac{\pi}{4}$ and calculate the CP asymmetry by using Eq.\eqref{CPN}. We show the variation of this coupling with the CP asymmetry in Fig \ref{cpn}, from which one can infer that for low values of the Yukawa couplings the CP asymmetry turns out to be very small. But considering a range of $0.02$ to $0.04$, one can achieve a large CP asymmetry of the order $10^{-7}-10^{-8}$, which is required for successful leptogenesis. Similarly, in Fig.\ref{cph}, we have shown the variation of triplet-lepton Yukawa coupling, constrained from the type II neutrino mass and out of equilibrium decay width of the scalar triplet. A favored region of the coupling $y^\prime_{ti} \mu_{iL}$ to vary within $1$ to 3 TeV for the decaying triplet and 100 to 200 TeV for the heavier triplet scalar to obtain a CP asymmetry of order $10^{-7}$.
\subsection*{Boltzman Equations}
Efficiency of leptogenesis plays a vital role in generating the final baryon asymmetry, which could be governed by the dynamics of relevant Boltzmann equations. The particle dynamics in the early universe indulge a large number of interactions in the thermal soup. Particles attain thermal equilibrium and are subjected to the chemical equilibrium constraints, while the gauge interaction rate is more than the Hubble expansion. Here the chemical potential becomes important to define the relations between the particles in the chemical equilibrium. For the interactions that are equal with the Hubble expansion are not that fast to remain in equilibrium. Therefore the Boltzmann equations are very much significant to analyze the particle number density after the chemical or kinetic decoupling of particles in a specific temperature regime. In this model, we consider two Higgs triplets, where one of the Higgs triplets is more massive than the other. Hence the asymmetry generated by the heavier triplet will be washed out by the inverse decay of lighter one. As it has been demonstrated in the literature earlier that the lepton number violation demands both the decay modes of the triplets to happen and any of the decay modes needs to be out of equilibrium to satisfy the Sakharov's condition. Unlike the right-handed neutrinos, as the scalar triplet is not Majorana particle and hence there will be asymmetry in particle and antiparticle decays and will contribute to the total asymmetry. 
\begin{eqnarray}
&& H(T)=\frac{4 {\pi}^3 g_\star}{45} \frac{T^2}{M_{pl}}, \hspace{3mm}  \text{where,} \hspace{3mm}    M_{pl}=1.2 \times 10^{19} \text{GeV},\\
&& Y^{eq}_{\Delta}= \frac{45   g_T}{4 {\pi}^4 g_\star} z^2 K_2(z), \hspace{3mm} Y^{eq}_N= \frac{135 \zeta{(3)} g_N}{16 {\pi}^4 g_\star} z^2 K_2(z),\\
&& {Y^{eq}_l}= \frac{3}{4} \frac{45 \zeta(3) g_l}{2 {\pi}^4 g_{\star}}, \hspace{3mm} {Y^{eq}_h}= \frac{45 \zeta(3) g_h}{2 {\pi}^4 g_\star},
\end{eqnarray}
where, $g_\star=106.75$ is the total relativistic degree of freedom of the SM particles in the equilibrium. $g_l=2$, $g_h=1$, $g_T=1$, $g_N=2$ are the degrees of freedom of lepton, Higgs doublets and Higgs triplet and right handed neutrinos, respectively, and $z=\frac{M_{\Delta_1}}{T}$, with $M_{\Delta_1}$ being the mass of decaying particle. The co-moving entropy density is given by $s=(\frac{2 {\pi}^2}{45}) g_\star T^3$, $\zeta(3) \approx 1.202$, and $K_i(z)$ are the modified Bessel functions of type i. Let us define the total co-moving number density of triplet be given by $\Sigma_\Delta =Y_{\Delta_i} + Y_{\bar{\Delta_i}}$, where, $Y_x=\frac{n_x}{s}$ are the co-moving number densities. The asymmetric densities of the particles are given by $\eta_x=Y_x-Y_{\bar{x}}$. Hence we can write the Boltzmann equations for different reaction densities to show the evolution of asymmetries $\eta_{\Delta}$, $\eta_h$, $\eta_l$, and are given in \cite{HahnWoernle:2009qn},\cite{Felipe:2013kk}:
\begin{eqnarray}
&& szH(z)\frac{d\Sigma_{\Delta_1}}{dz}=-\left(\frac{\Sigma_{\Delta_1}}{{\Sigma^{eq}_{\Delta_1}}}-1\right)\gamma_D-2 \left(\frac{{\Sigma^{2}_{\Delta_1}}}{{{\Sigma^{eq}_{\Delta_1}}}^2}-1\right)\gamma_A, \\
&& szH(z)\frac{d\eta_{\Delta_1}}{dz}=-\gamma_D\left(\frac{\eta_{\Delta_1}}{{\Sigma^{eq}_{\Delta_1}}}+\sum B_l \frac{\eta_{l}}{{Y^{eq}_l}}-B_h \frac{\eta_h}{{Y^{eq}_h}}\right),\\ 
&& szH(z)\frac{d\eta_h}{dz}=2\gamma_D\left(\sum B_l \epsilon_l-B_h \epsilon_h\right)-2B_h \gamma_D\left(\frac{\eta_h}{{Y^{eq}_{h}}}-\frac{\eta_{\Delta_1}}{{\Sigma^{eq}_{\Delta_1}}}\right), \\
&& szH(z)\frac{d\eta_l}{dz}= 2\gamma_D\left(\sum B_l \epsilon_l-B_h \epsilon_h \right)-2B_l \gamma_D\left(\frac{\eta_l}{{Y^{eq}_l}}+\frac{\eta_{\Delta_1}}{{\Sigma^{eq}_{\Delta_1}}}\right).
\end{eqnarray}
The decays and inverse decays are important in the Boltzmann equations to contribute to the lepton asymmetry, which are given by
\begin{eqnarray}
\gamma_D=s \Gamma_{\Delta_1} \Sigma_{\Delta_1} \frac{K_1(z)}{K_2(z)}, 
\end{eqnarray}
and the s-wave contribution to the gauge scattering processes of triplets is 
\begin{equation}
\gamma_A=\frac{M_{\Delta_1} T^3 e^{\frac{-2 M_{\Delta_1}}{T}}}{64 \pi^4}(9 g^4+ 12 g^2 g^2_1 + 3 g^4_1) \left(1+ \frac{3 T}{4 M_{\Delta_1}}\right),
\end{equation} 
where, g and $g_1$ are the SM gauge couplings. Along with the gauge scattering for triplets, we can also have $\Delta L=2$ , lepton number violating interactions. But we can safely neglect them in this model, as those will be suppressed by the heavy mass of  mediating particle($\Delta_1$).\\
\begin{figure}
\includegraphics[height=60mm,width=75mm]{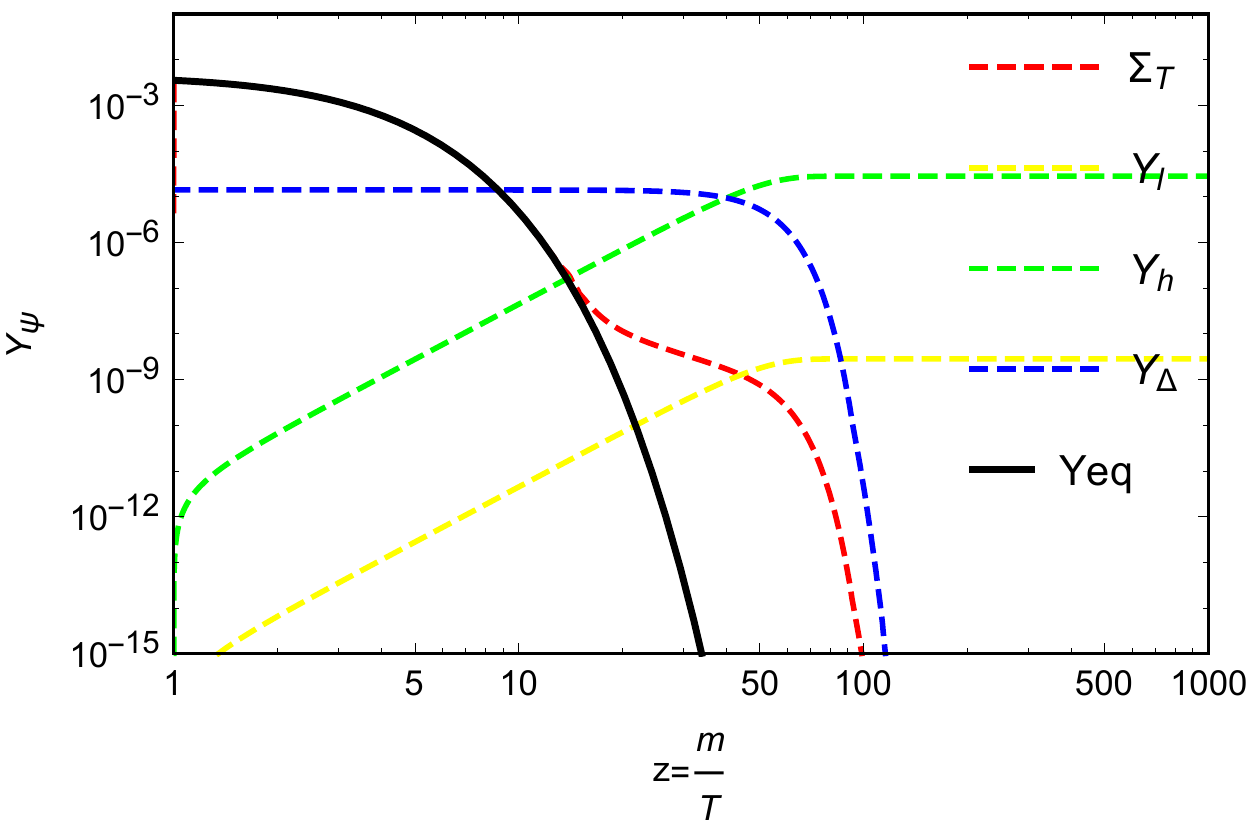}
\caption{ Figure shows the abundances of particles when lepton asymmetry is generated solely from scalar triplet. Even though the lepton asymmetry from the lepton loop vanishes, still a large asymmetry can be generated from scalar and right handed neutrino loop.}
\label{beqn}
\end{figure}
Hence from Fig \ref{beqn} one can see that a lepton asymmetry of order $\approx 10^{-9}$ can be achieved by using the Boltzmann equation in high mass regime of the scalar triplet. This can be converted to the baryon asymmetry during sphaleron transition with a fraction of $Y_B= a Y_L$, where, $a=- \frac{8 N_F +4 N_H}{22 N_F +13 N_H} \approx -0.34$. $N_H$ and $N_F$ are the number of Higgs and fermion generations, respectively.
\subsection{Low scale Leptogenesis: $M_\Delta \approx \mathcal{O}(2)$ TeV}
High scale leptogenesis with such a heavy triplet is very difficult to have any experimental signature in the near future. But efforts have been made to bring down the scale of leptogenesis with resonance effect, which can also explain the current neutrino oscillation data \cite{Pilaftsis:2003gt}-\cite{Abada:2018oly}. This low scale is not only phenomenologically viable but also can be verified by the collider experiments. We discuss the impact of leptogenesis in TeV scale along-with the constraints on coupling from the neutrino masses. We also put light on the contribution of TeV scale triplets and right-handed neutrinos to the muon $g-2$ anomaly and  lepton flavor violating (LFV) rare decays.\\
\textbf{Constraint on couplings from out of equilibrium decay and neutrino mass:}\\
To generate nonzero asymmetry as per the Sakharov's conditions the decay rate should be less than the Hubble expansion 
\begin{eqnarray}
\Gamma_{\Delta_1} & < & H \approx 1.66 \times \sqrt{g_\star}\frac{T^2}{1.2\times 10^{19}}~ \rm GeV \nn \\
&& < 1.38\times 10^{-18} [T^2]_{T=M_{\Delta_1}} (g_\star \approx 100)~ \rm GeV.
\end{eqnarray}
From type I:
\begin{eqnarray}
&& \sum m^I_\nu = \frac{4y^2_{\nu_1} v^2 \sin^2{\beta}}{M_{1R}}+\frac{2 y^2_{\nu_3} v^2 \sin^2{\beta}}{M_{1R}}+\frac{2 y^2_{\nu_4} v^2 \cos^2{\beta}}{M_{1R}}\leq 0.05\times 10^{-9}~ \rm GeV \nn \\
&& \approx 0.3 y^2_{\nu_3}\sin^2{\beta}+y^2_{\nu_4} \cos^2{\beta} \leq 4.1\times 10^{-12}.
\end{eqnarray}
From type II:
\begin{eqnarray}
&& \sum_i m^{II}_\nu = \frac{(4y_{ti}+2{y_{ti}}')\mu_{iL} v^2}{{M_{\Delta i}}^2} \leq 0.05\times 10^{-9}~ \rm GeV \nn \\
&&\approx (2 y_{ti}+{y_{ti}}')\mu_{iL} \leq 8\times 10^{-10}~ \rm GeV~ (\rm for\hspace{2mm} \rm i=1), \nn \\
&& \leq 8.3 \times 10^{-8}~\rm GeV~ (\rm for \hspace{2mm} \rm i=2).
\end{eqnarray}

As the CP asymmetry turns out to be very small, another way to realize the leptogenesis is through resonance enhancement. If both of the triplets will be quasi degenerate in mass, there will be large enhancement in CP asymmetry. The expression of CP asymmetry can be written as \cite{Hambye:2012fh}
\begin{equation}
\epsilon_{CP}=\frac{1}{2 \pi} \frac{\rm Im\left[Y_{\Delta_1} {Y^\dagger_{\Delta_2}}\mu^\star_{\Delta_1}\mu_{\Delta_2}\right]}{M^2_{\Delta_1} (Y^\dagger_{\Delta_1} Y_{\Delta_1})_{ii}+\mu^2_{\Delta_1}} \frac{M^2_{\Delta_1} \delta M^2_{12}}{(\delta M^2_{12})^2+M^2_{\Delta_1} \Gamma^2_{\Delta_2}}. 
\end{equation}
Where, $\delta {M^2_{12}}={M^2_{\Delta_2}}-{M^2_{\Delta_1}}$. With the standard assumption of resonance case, provided in the literature \cite{Pilaftsis:2003gt}, $\delta M^2_{12} \approx M_{\Delta_1} \Gamma_{\Delta_2} \ll M^2_{\Delta_1}$, we can have the self energy contribution
\begin{equation}
\frac{M^2_{\Delta_1} \delta {M^2_{12}}}{(\delta M^2_{12})^2+M^2_{\Delta_1} \Gamma^2_{\Delta_2}} \approx \frac{M_{\Delta_1}}{\Gamma_{\Delta_2}}\gg 1.
\end{equation}
Hence there will be resonance enhancement in the self energy,  which contributes maximally to the CP asymmetry. With such a large CP asymmetry ($\approx \mathcal{O}(1)$), the observed baryon asymmetry can be explained by solving the Boltzmann equations as in the previous case. But at low energy the washout processes will dominantly contribute and the desired lepton asymmetry can be achieved.
\begin{table}[h]
\begin{center}
\begin{tabular}{| c | c | c | c | c | c| c| c|}
\hline
~Parameters~ & $M_{\Delta_1}$(TeV) & $M_{\Delta_2}$(TeV) & ${y_{t1}}'{\mu_{1L}}$ (GeV) & ${y_{t2}}'{\mu_{2L}}$ (GeV) &~ $\sum{m_\nu}$(eV) ~& ~$\epsilon_{CP}$ \\
\hline
 BP1 & $2$ & $20$ & $7.2\times 10^{-10}$ & $9\times 10^{-7}$ & $0.023$ & 0.06\\
 \hline
BP1 & $2$ & $20$ & $6\times 10^{-10}$ & $7.6\times 10^{-7}$ & $0.02$ & 0.1\\
   \hline
\end{tabular}
\end{center}
\caption{Benchmark points for the parameters satisfying the constraints from neutrino mass and observed baryon asymmetry.}
\end{table}
\subsubsection{\textbf{Comments on LFV Decays and muon g-2 anomalies}}
\begin{figure}
\includegraphics[height=30mm,width=50mm]{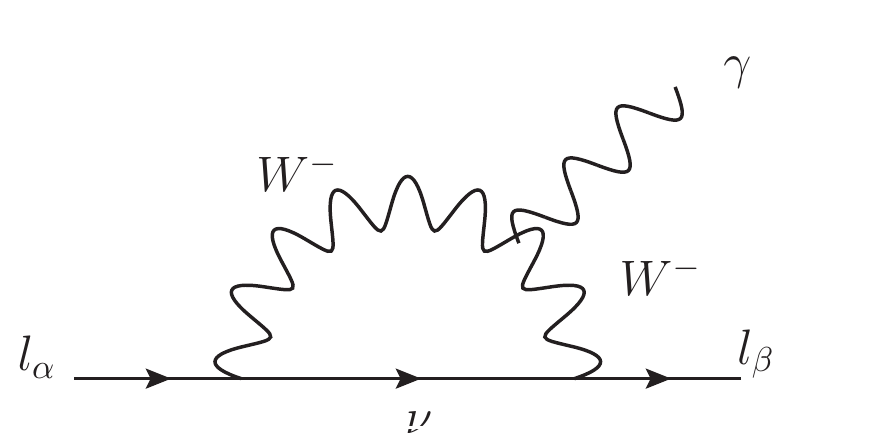}
\includegraphics[height=30mm,width=50mm]{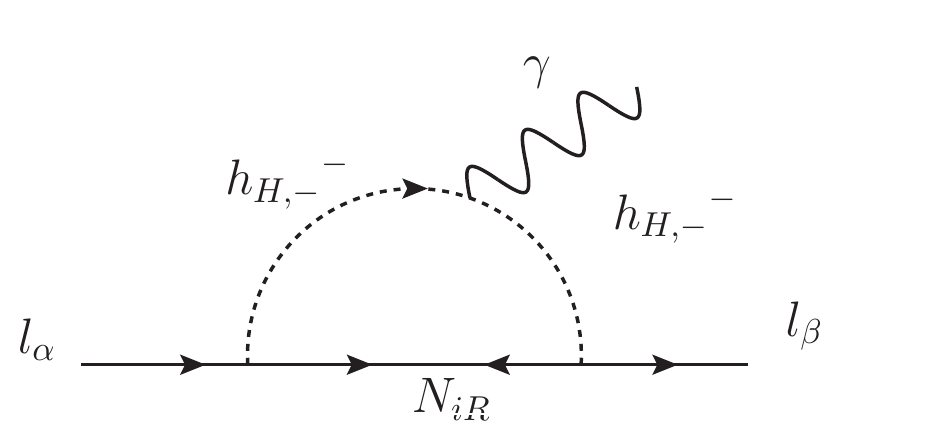}
\includegraphics[height=30mm,width=50mm]{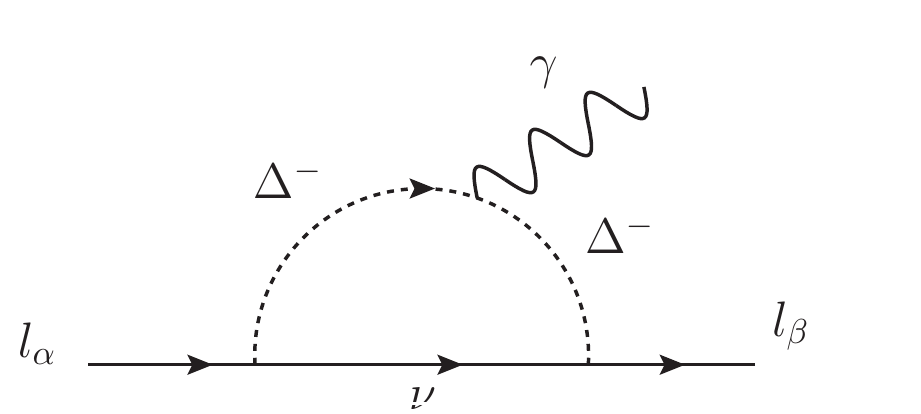}
\caption{Feynman diagrams represent the Lepton flavor violating rare decays and muon g-2 anomaly in one loop level.}\label{lfvfeyn}
\end{figure}
Lepton flavor violating decay processes have received great attention in recent times which are very rare to be observed experimentally \cite{Mihara:2013zna}-\cite{Dev:2017ftk}. Efforts are being made by many experiments to look in this direction and some of them have provided a stringent upper limits on these decays. In this context, $\mu \rightarrow e\gamma$ looks to be an important process to be measured with less background from observation point of view. The current experimental limit on this decay is Br$(\mu\rightarrow e\gamma)<4.2\times 10^{-13}$ from MEG collaboration \cite{TheMEG:2016wtm}. In the framework of low scale leptogenesis, we can have extra contribution to rare decays $l_\alpha\rightarrow l_\beta \gamma$ due to the presence of right handed neutrinos and Higgs. Due to the diagonal structure of scalar triplet-Yukawa coupling, there won't be any contribution to LFV from triplet sector but we can still have it from right handed neutrino and heavy Higgs loop. The branching ratio for this decay is given by \cite{Chekkal:2017eka}
\begin{equation}
Br(l_\alpha \rightarrow l_\beta \gamma)=\frac{3(4 \pi)^3 \alpha}{4 {G_F}^2}|A_D|^2\times Br(l_\alpha \rightarrow l_\beta \nu_\alpha \bar{\nu_{\beta}}).
\end{equation}
where, $G_F\approx 10^{-5}{GeV}^{-2}$ is the Fermi constant and $\alpha$ is the electromagnetic fine structure constant and $A_D$ is the dipole contribution, which is given by
\begin{equation}
A_D=\sum_i \frac{(Y^\nu_{H,-})_{\alpha i}(Y^{\nu \star}_{H,-})_{\beta i}f(x)}{2(4\pi)^2 M^2_{h_H}}.
\end{equation} 
Here $Y^\nu_{H,-}$'s are the Yukawa coupling matrices in Eq.\eqref{yukawaN}, $M_{h_H}$ is the mass of heavy Higgs ($H_H,H_-$) and $f(x)$ is the loop function, with $x=\frac{M^2_{iR}}{M^2_{h_H}}$, and is given by
\begin{equation}
f(x)=\frac{1-6x+3x^2+2x^3-6x^2 {\rm log}x}{6(1-x)^4}.
\end{equation} 
When we have $\alpha=\beta$, the above diagrams will contribute to the muon anomalous magnetic moment as
\begin{equation}
\delta a_\mu=\frac{1}{16 \pi^2}\left[\frac{m^2_\mu}{M^2_{h_H}}\sum_i |(Y^\nu_{H,-})_{\mu \mu}|^2f(x)+\frac{m^2_\mu}{M^2_{\Delta_i}}\sum_i |y_{ti}|^2f(x_t)\right],
\end{equation}
where, $f(x_t)$ will follow the same expression of $f(x)$, with $x_t=\frac{m^2_\nu}{M^2_{\Delta_i}}$.\\
\begin{figure}
\includegraphics[height=40mm,width=50mm]{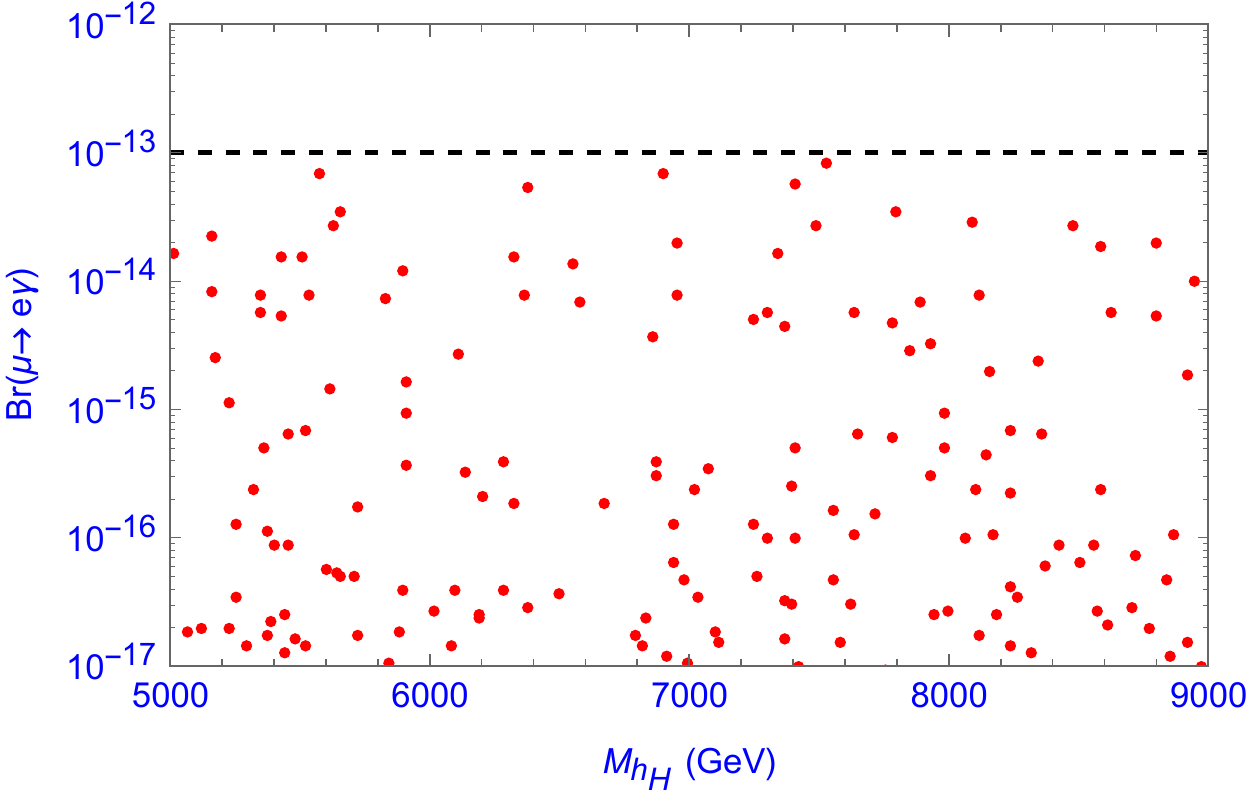}
\includegraphics[height=40mm,width=50mm]{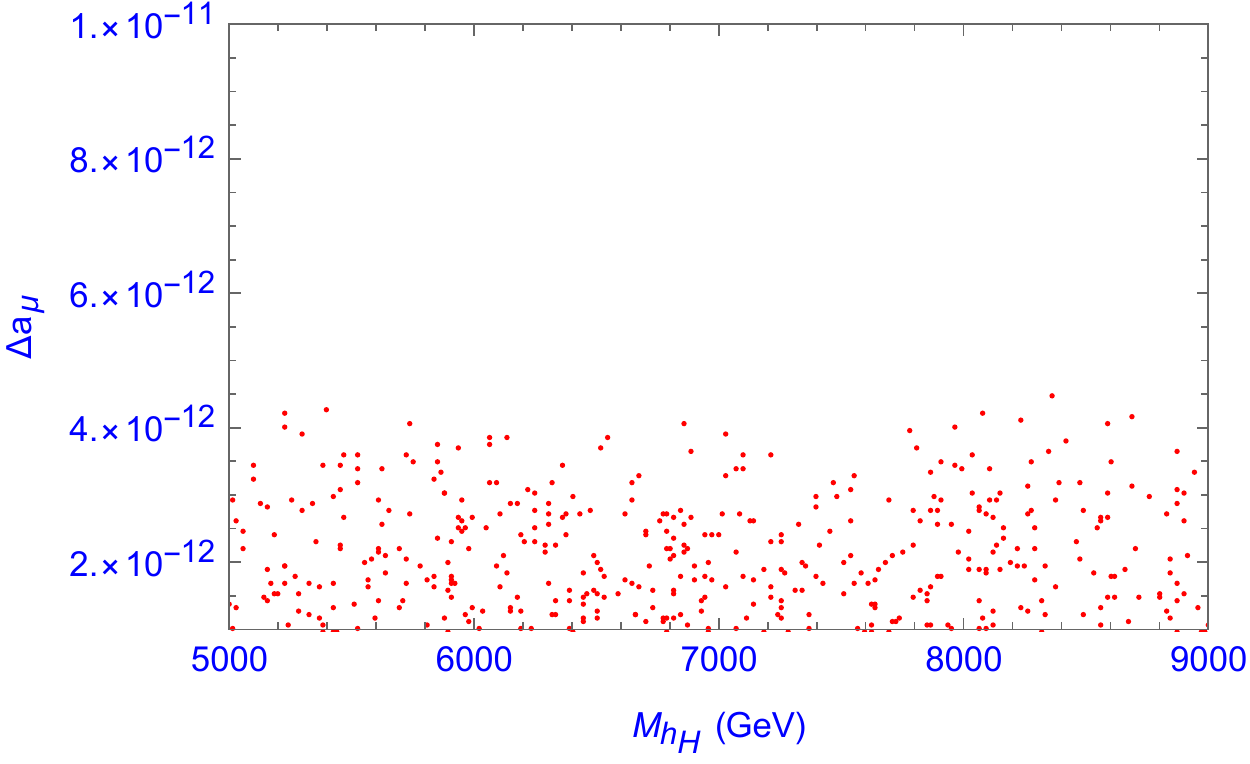}
\includegraphics[height=40mm,width=50mm]{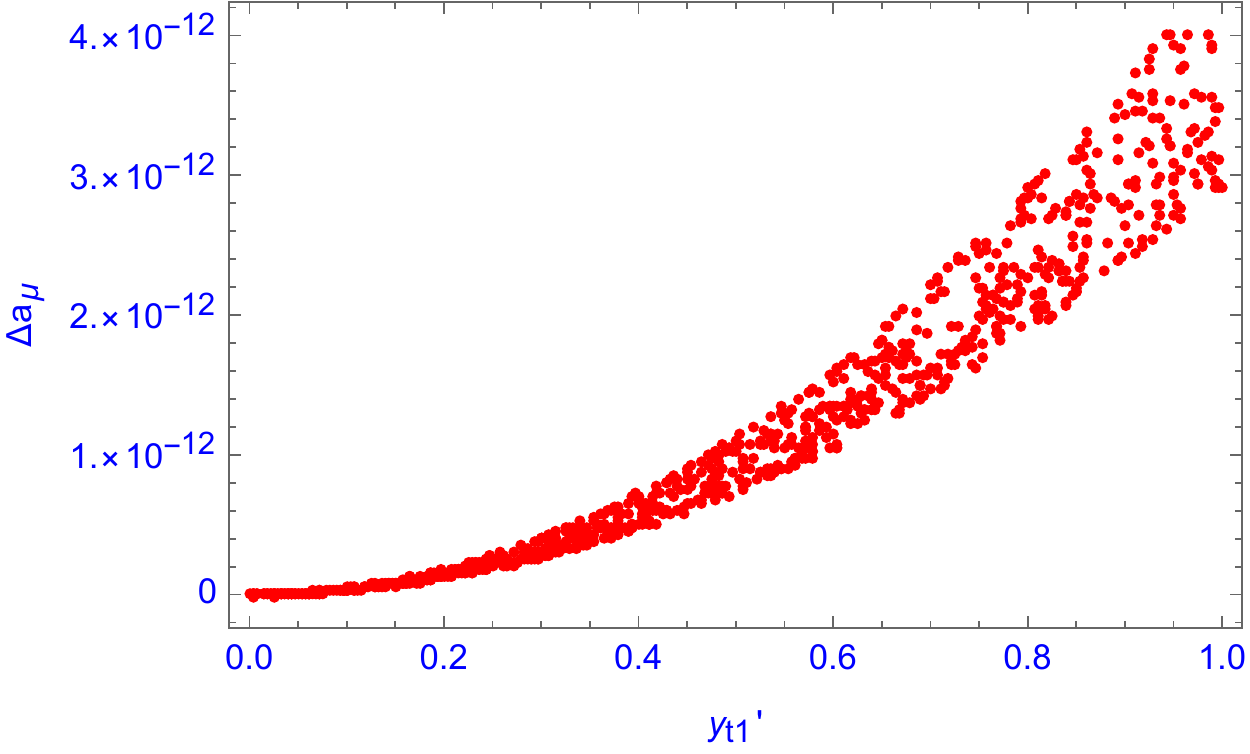}
\caption{The left most panel shows the parameter space of heavy Higgs mass allowed by the branching of lepton flavor violating decay $\mu \rightarrow e \gamma$, which is coming in order of less than $10^{-13}$ as per the experimental bound. The middle panel shows the variation of same Higgs mass that gives the viable range of muon anomalous magnetic moment allowed by experiment and the right most panel represents the variation of triplet-lepton Yukawa with muon anomalous magnetic moment.}
\label{lfvplots}
\end{figure}
From the left panel of Fig.\ref{lfvplots}, one can see that there will be extra contribution to rare LFV decay $\mu \rightarrow e \gamma$ from the right-handed neutrinos and heavy Higgs mediated diagram (shown in the middle panel of Fig.\ref{lfvfeyn}) with a mass scale of order $\mathcal O(10)$ TeV. Middle panel of the Fig.\ref{lfvplots} represents the appreciable contribution to muon g-2 anomaly due to the presence of extra particles in the model shown in the middle and right panels of Fig.\ref{lfvfeyn}. The extreme right panel of the Fig.\ref{lfvplots} displays the variation of the triplet Yukawa coupling with $\Delta a_\mu$, where the $\Delta a_\mu$ is found to be order of $\mathcal O(10^{-12})$ by varying the Yukawa coupling from 0 to 1 and the triplet mass of the order $\mathcal O (1.6)$ TeV.
\section{Summary}
In this article, an attempt has been made to understand the lepton asymmetry using the simplest discrete symmetry $S_3$  along with the $Z_2$ symmetry. Here, we first considered the type-II seesaw mechanism by introducing the scalar triplets but the triplet-lepton Yukawa matrix turned out to be diagonal. So in order to explain the neutrino masses and mixing we included additional right-handed neutrinos in the type-I seesaw scenario. We considered the combination of both type-I and type-II seesaw scenarios to accommodate current results in the neutrino sector, so also attempted to explain the observed baryon asymmetry. Due to the presence of both right-handed neutrinos and scalar triplets we considered two different scenarios, in which the first case we discussed the scenario where the scalar triplets are lighter than the right-handed neutrino masses and explained the lepton asymmetry in high mass regime. Thereafter, we considered the case where the masses of RHNs and scalar triplets are much smaller (of the order of TeV), which can be tested in future colliders, and explained the leptogenesis by resonance enhancement from self energy loop with quasi degenerate triplets. We used the Boltzmann equation for the first case and discussed the results which showed that the combined effect of type-I+II enhances the CP asymmetry. There is no contribution from the triplet mediated process but due to the presence of right-handed neutrinos and heavy Higgs there will be contributions to the $\mu \rightarrow e \gamma$ process and hence one can explain the available LFV result. Moreover, in this model for the muon anomalous magnetic moment, there will be contributions from the type I and II sectors. Hence the current framework is more interesting as it not only explains the neutrino mass and leptogenesis results but also satisfies the experimental bounds on lepton flavor violating branching ratio and muon anomalous magnetic moment simultaneously in the low mass regime with quasi degenerate triplets. Interestingly, the low mass regime also opens up the exciting possibilities of scalar triplets to be tested in future collider experiments.

Subhasmita Mishra would like to thank DST Inspire for the financial support. The authors would also like to thank Narendra Sahu for useful discussions.
\section{Appendix}
 $S_3$ group includes 3 dimensional reducible representation which can be reduced to a doublet and a singlet (i.e., $3 = 2 \oplus 1 $). If ($a_1\hspace{3mm}a_2$) and ($b_1 \hspace{3mm} b_2$) transform as doublets under $S_3$, the tensor product rules are summarized as  \cite{Haba:2005ds},
\begin{eqnarray*}
&& \left(a_1 \hspace{2mm} a_2\right)^T_2\otimes \left(b_1 \hspace{2mm} b_2\right)^T_2=\begin{pmatrix}
a_1 b_2+a_2 b_1\\
a_1 b_1-a_2 b_2\\
\end{pmatrix}_2\oplus \left(a_1 b_1+a_2 b_2\right)_1 \oplus \left(a_1 b_2-a_2 b_1\right)_{1^{\prime}}.\\ \nonumber
\\ 
&& \left(a_1 \hspace{2mm} a_2\right)^T_2\otimes (b^{\prime})_1=\left(a_1 b^{\prime} \hspace{2mm} a_2 b^{\prime} \right)^T_2, \hspace{5mm} \left(a_1 \hspace{2mm} a_2\right)^T_2 \otimes (b^{\prime})_{1^{\prime}}=\left(-a_2 b^{\prime} \hspace{2mm} a_1 b^{\prime} \right)^T_2.\\
&& (a)_1 \otimes (b)_{1^{\prime}}=(ab)_{1^{\prime} },\hspace{6mm} (a)_{1^{\prime}}\otimes (b)_{1^{\prime}}=(ab)_1.
\end{eqnarray*}



\end{document}